\documentclass[10pt,journal,doublecolumn]{IEEEtran} %,compsoc
\IEEEoverridecommandlockouts
\usepackage{amsfonts, amsmath,amssymb,amsthm}
\usepackage{algorithm, algorithmic}
\usepackage{array}
\usepackage{cases}
\usepackage{changepage}
\usepackage[compress]{cite}
\usepackage[english]{babel}
\usepackage{etoolbox}
\usepackage{graphicx}   % includegraphics
\usepackage{grfext}     % lets us add extra extensions for graphicx
\usepackage{epstopdf}   % defines conversion rule eps->pdf (needs shell-escape)

\AppendGraphicsExtensions{.eps}
\usepackage{multirow}
\usepackage{physics} 	%	For partial derivative
\usepackage{subfigure}
\usepackage{setspace}
\usepackage{url}  % This makes \url work
\usepackage{lipsum}
\usepackage{nicefrac}	% better inline fraction mode
\usepackage[table,xcdraw]{xcolor}	%Table color
\usepackage{textcomp}% to input temperature unit degrees Celsius
\usepackage{makecell}
\usepackage[font=small]{caption} %	Configure figure caption font
\usepackage{booktabs} % For table formatting
\usepackage{mathrsfs}% Flower body
\usepackage{indentfirst} % Add indent to the first paragraph
\usepackage{textcomp} % Generate degree symbol (textdegree)
\usepackage{pifont}		% To use dingbat symbols
\usepackage{threeparttable}
\usepackage[right]{lineno}  % switch option ensures numbers are on the outer margin
\def\ps@headings{%
\def\@oddhead{\mbox{}\scriptsize\rightmark \hfil \thepage}%
\def\@evenhead{\scriptsize\thepage \hfil \leftmark\mbox{}}%
\def\@oddfoot{}%
\def\@evenfoot{}}
\makeatother
\makeatletter
\patchcmd{\@thm}
  {\trivlist}
  {\list{}{\leftmargin=\thm@leftmargin\rightmargin=\thm@rightmargin}}
  {}{}
\patchcmd{\@endtheorem}
  {\endtrivlist}
  {\endlist}
  {}{}
\newlength{\thm@leftmargin}
\newlength{\thm@rightmargin}

\newcommand{\xnewtheorem}[3]{%
  \newenvironment{#3}
    {\thm@leftmargin=#1\relax\thm@rightmargin=#2\relax\begin{#3INNER}}
    {\end{#3INNER}}%
  \newtheorem{#3INNER}%
}

\makeatother

\newtheoremstyle{indentedupright}{3pt}{3pt}{} {}{\bfseries}{.}{.5em}{} % Style 1
\newtheoremstyle{indenteditalic}{3pt}{3pt}{\itshape} {}{\bfseries}{.}{.5em}{} % Style 2

\theoremstyle{indenteditalic}
\xnewtheorem{0pt}{0pt}{oldpro}{Conventional Property}
\xnewtheorem{0pt}{0pt}{newpro}{New Property}
\xnewtheorem{0pt}{0pt}{obser}{Observation}
\xnewtheorem{0pt}{0pt}{pro}{Property}
\xnewtheorem{0pt}{0pt}{lem}{Lemma}
\xnewtheorem{0pt}{0pt}{corl}{Corollary}
\xnewtheorem{0pt}{0pt}{addcon}{Additional Constraint}

\newcommand{\romu}[1]{\uppercase\expandafter{\romannumeral #1\relax}} % Upper case of roman number, e.g., Rom{1}
\newcommand{\roml}[1]{\lowercase\expandafter{\romannumeral #1\relax}}    % Lower case of roman number, e.g., rom{1} 

\renewcommand{\baselinestretch}{1}

\begin{document}
\title{\LARGE Experimental Study of Underwater Acoustic Reconfigurable Intelligent Surfaces with Synthetic Reflection}
\author{\IEEEauthorblockN{Yu Luo\IEEEauthorrefmark{1}, Lina Pu\IEEEauthorrefmark{2}}, Aijun Song\IEEEauthorrefmark{3}\\
\IEEEauthorblockA{\IEEEauthorrefmark{1}ECE Department, Mississippi State University, Mississippi State, MS, 39759, USA\\
\IEEEauthorrefmark{2}Department of Computer Science, University of Alabama, Tuscaloosa, AL 35487, USA\\
\IEEEauthorrefmark{3}Department of Electrical and Computer Engineering, University of Alabama, Tuscaloosa, AL 35487, USA\\
Email: yu.luo@ece.msstate.edu, lina.pu@ua.edu, song@eng.ua.edu}
}

%\linenumbers
\maketitle
\begin{abstract}
\label{:Abstract}
This paper presents an underwater acoustic reconfigurable intelligent surface (UA-RIS) designed for long-range, high-speed, and environmentally friendly communication in oceanic environments. The proposed UA-RIS comprises multiple pairs of acoustic reflectors that utilize a synthetic reflection scheme to flexibly control the amplitude and phase of reflected waves. This capability enables precise beam steering to enhance or attenuate sound levels in specific directions. A prototype UA-RIS with 4$\times$6 acoustic reflection units is constructed and tested in both tank and lake environments to evaluate performance. Experimental results using a continuous wave (CW) as the source signal demonstrate that the prototype is capable of effectively pointing reflected waves to targeted directions while minimizing side lobes through synthetic reflection. Field tests reveal that deploying the UA-RIS on the sender side considerably extends communication ranges by 28\% in deep water and 46\% in shallow waters. Furthermore, with a fixed communication distance, positioning the UA-RIS at the transmitter side substantially boosts the receiving signal-to-noise ratio (SNR), with an average increase of 2.13\,dB and peaks up to 2.92\,dB. When positioned on the receiver side, the UA-RIS can expand the communication range in shallow and deep water environments by 40.6\% and 66\%, respectively. Moreover, placing the UA-RIS close to the receiver enhances SNR by an average of 2.56\,dB, reaching up to 4.2\,dB under certain circumstances. 

\end{abstract}


%%=========================================
%%=========================================

\section{Introduction}
\label{sec:Intr}
In recent years, radio frequency reconfigurable intelligent surfaces (RF-RIS) have emerged as a enabling technology to enhance communication quality for 5G and beyond networks~\cite{yang2023beyond}. By dynamically adjusting the amplitude and phase of reflected electromagnetic waves with the load network, RIS can precisely steer beams in specific directions. This capability enhances signal strength at the receiver while reducing interference in surrounding environments, thereby improving the overall performance of wireless communication networks~\cite{wu2024intelligent}. 

Extensive research has been conducted for RIS in radio environments, bringing this cutting-edge technology into underwater environments remains a complex problem. In water, the transmission range of electromagnetic (EM) waves is severely restricted due to significant absorption attenuation. Currently, acoustic communication is still the principal method for mid-range and long-range communication in marine settings. However, the physical properties of acoustic waves are completely distinct from those of radio signals, rendering existing RF-RIS unsuitable for direct application in water. At present, only limited studies explore the design of underwater acoustic RIS (UA-RIS)~\cite{sun2022high, wang2023designing}.

\begin{figure}[htb]
\centerline{\includegraphics[width=8.5cm]{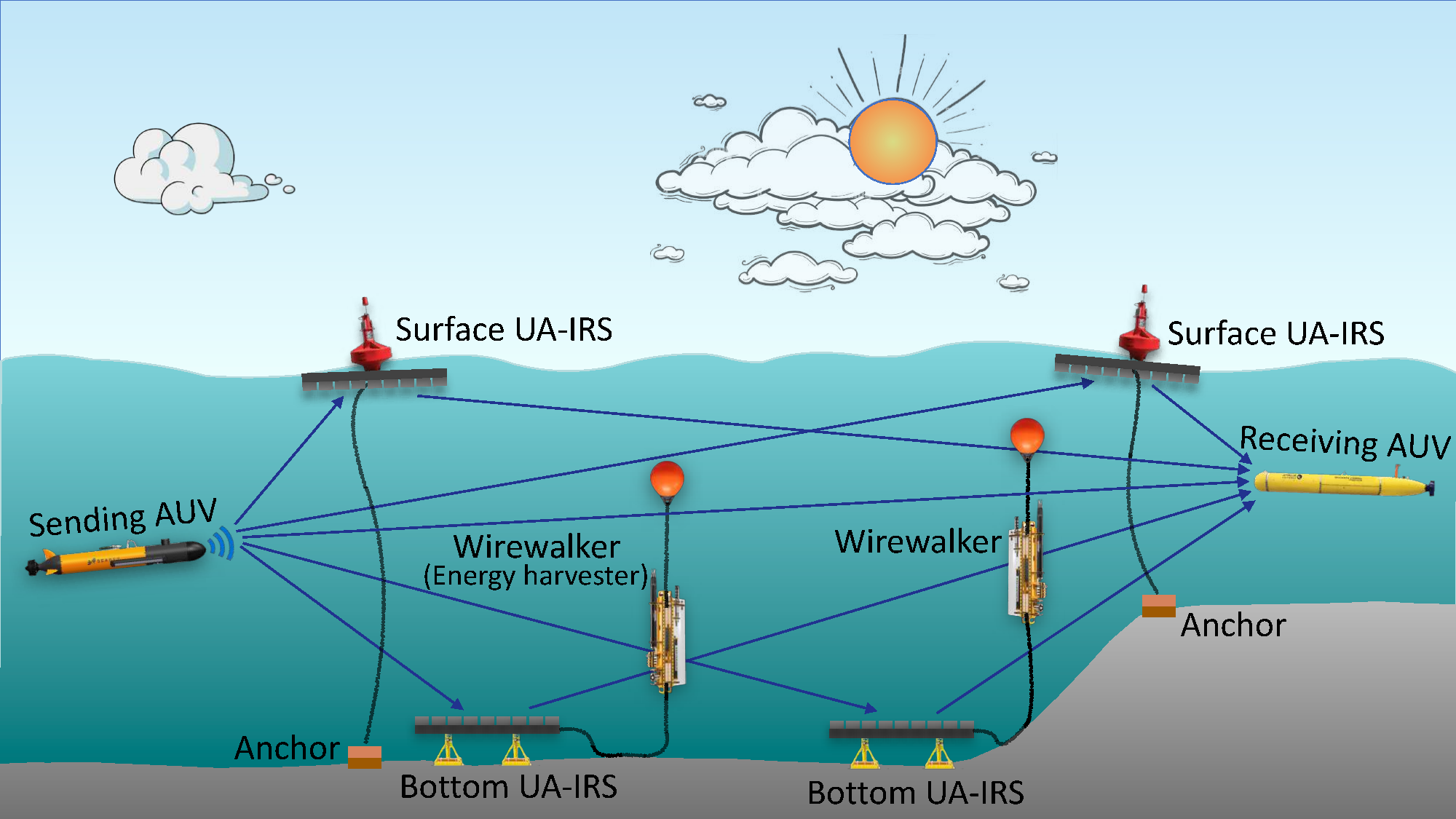}}
  \caption{UA-RIS assisted long-range underwater acoustic communication network.}\label{fig:Scenario}
\end{figure}

\textcolor{black}{UA-RIS can significantly extend the communication range, making it a promising solution for achieving ultra-long-range underwater communication.} Traditional acoustic systems typically operate at low frequencies to reduce propagation loss, but this often requires either bulky transducers or higher transmission power, both of which can reduce battery life. In contrast, UA-RIS enables the use of compact, low-power transducers while still maintaining effective long-range performance \textcolor{black}{by strategically deploying multiple intelligent surfaces at both the ocean surface and seabed}. As illustrated in Fig.~\!\ref{fig:Scenario}, the surface UA-RIS unit is solar-powered and mounted on a buoy, while the deep-sea UA-RIS is connected to a wirewalker~\cite{del2021wirewalker}, a fully passive device that harvests energy from the motion of ocean waves.

\textcolor{black}{Through careful design, UA-RIS units can precisely direct the source signal toward the intended receiver, coherently combining reflected acoustic waves to significantly enhance the signal-to-noise ratio (SNR). This enables extended communication ranges, even at higher frequencies, by effectively preserving signal strength over long distances.}

\begin{figure}[htb]
\centerline{\includegraphics[width=9cm]{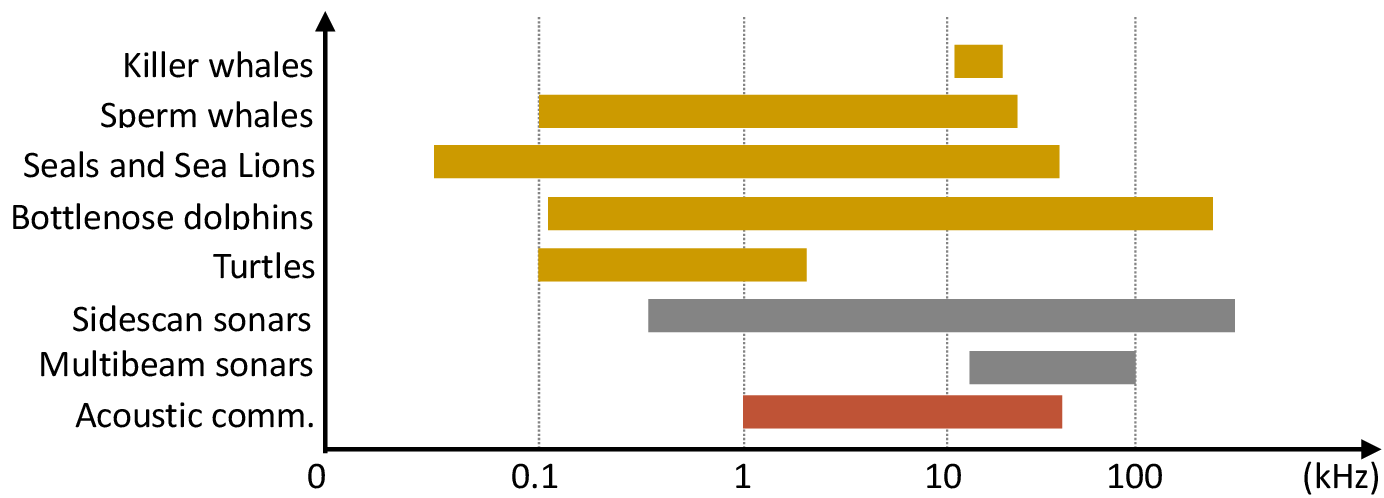}}
  \caption{Typical channel usage of marine life and human-made acoustic systems.}\label{fig:SpecShar}  
\end{figure}

\textcolor{black}{In addition to extending the communication distance, UA-RIS can reduce the source level of acoustic signals without compromising communication quality, enabling more environmentally friendly underwater communications.} Marine wildlife, such as dolphins and whales, rely heavily on sound for communication, hunting, and navigation~\cite{luo2014challenges}. As shown in Fig.~\!\ref{fig:SpecShar}, bottlenose dolphins use frequencies from 200\,Hz to 24\,kHz for whistles and 200\,Hz to 150\,kHz for echolocation clicks, while killer whales' echolocation typically occurs within the 12\,kHz to 25\,kHz range~\cite{richardson2013marine}. These ranges overlap with those of human-made acoustic systems, potentially causing interference or auditory harm.

\textcolor{black}{UA-RIS technology provides a viable solution to the above challenge. By enabling acoustic modems to transmit at lower sound levels, it reduces environmental impact and minimizes interference with the acoustic channel. Additionally, UA-RIS focuses reflected signals in specific directions rather than dispersing them, thereby reducing interference and enhancing spatial channel utilization.}

\textcolor{black}{Although UA-RIS offers advantages in long-range and environmentally friendly communication, transitioning this technology from theory to practical application requires overcoming several key challenges:}

\vspace{0.1cm}
\begin{adjustwidth}{-0.71cm}{0cm}
\begin{description}
\setlength{\labelsep}{-0.95em}
\itemsep 0.07cm
  \item[a)] \textbf{Constrained energy supply:} UA-RIS systems operate in maritime settings, where stable power infrastructure is often absent. Consequently, UA-RIS units, especially those situated on the ocean floor, require energy harvesting from sources such as ocean waves or microorganisms for self-sufficiency~\cite{kelly2023prototyping}. Nonetheless, the power density provided by these sustainable sources tends to be low, presenting substantial challenges for UA-RIS design. Addressing this limitation necessitates meticulous planning in both system design and power management strategy.
  \item[b)] \textbf{High dynamic range of acoustic signal:} To mitigate the high propagation attenuation over long range, acoustic modems are designed with high source levels. As a result, the intensity of sound waves reaching the UA-RIS varies widely with the distance. This high dynamic strength of incident waves poses a substantial challenge for system development. The reflection unit must be able to handle both weak and strong signals, requiring versatile load network configurations to accurately reflect waves with the intended amplitude and phase.
  \item[c)] \textbf{Low acoustic frequency:} The frequency employed for mid-range and long-range acoustic communication is significantly lower than that of RF systems due to the substantial frequency-dependent absorption of acoustic waves in water. Consequently, the wavelength of acoustic signals becomes long. In this context, the phase of the reflected waves cannot be adjusted by changing the length of the transmission line, a method commonly applied in RF-RIS. UA-RIS necessitates more efficient phase-shift mechanisms to effectively control wave reflection.
\end{description}
\end{adjustwidth}
\vspace{0.1cm}

To address the above issues, we proposed a \textcolor{black}{new UA-RIS architecture}, which automatically pair acoustic reflectors based on the angles of reflected waves. Each pair of reflectors then performs \textcolor{black}{synthetic reflection} scheme to direct the reflected beam toward specific directions. Compared to the 1-bit and 2-bit coding schemes widely used in RF-RIS for beam steering~\cite{kim2023rotated, yang2023beyond}, the \textcolor{black}{synthetic reflection} offers significantly enhanced flexibility. Specifically, traditional multi-bit coding schemes can produce only a finite number of phase states. This limitation results in coarse control over the reflected waves, leading to less precise beam direction. Furthermore, multi-bit coding schemes typically do not allow altering the amplitude of the reflected wave, resulting in higher side lobes in the beam pattern, which can cause unexpected interference issues. In contrast, \textcolor{black}{synthetic reflection} enables the formation of reflected waves with arbitrary amplitude and phase, supporting more sophisticated array processing methods such as minimum variance distortionless response (MVDR) and linearly constrained minimum variance (LCMV)~\cite{van2002optimum}. As a result, the reflected waves can be accurately directed in the desired directions without introducing strong interferences in environments. 

A prototype equipped with 24 (6 $\!\!\times\!\!$ 4) acoustic reflectors is implemented to assess the effectiveness of the proposed UA-RIS. Each reflector is constructed with a Tonpilz structure to reflect incident waves efficiently at 28\,kHz frequency. The complete assembly, which includes the matching circuit and the load network, is controlled by an ultra-low-power microcontroller (MCU) and multiple I/O expanders. The whole system has been carefully optimized to reliably manage high-dynamic acoustic signals originating from various acoustic facilities and mobile platforms.

We carried out comprehensive \textcolor{black}{simulations using COMSOL Multiphysics}, along with real-world experiments in both tank and lake environments, to evaluate the performance of the proposed work across various configurations. The experimental results \textcolor{black}{using a CW signal as the source} show that the UA-RIS can flexibly enhance or attenuate the strength of acoustic waves in specific directions. By deploying the UA-RIS with 24 reflection units at the sender side, the receiving \textcolor{black}{SNR of acoustic communication can be improved by an average of 2.13\,dB (up to 2.92\,dB)}, or the communication range can be extended by 28\% and 46\% in deep water and shadow water environments, respectively. When positioned near the receiver, the UA-RIS extends the communication range in shallow and deep water environments by 40.6\% and 66\%, respectively. Additionally, situating the UA-RIS adjacent to the receiver boosts \textcolor{black}{SNR an average of 2.56\,dB, with increases up to 4.2\,dB under certain conditions}.

To summarize, our work makes three key contributions:
\vspace{0.1cm}
\begin{adjustwidth}{-0.71cm}{0cm}
\begin{description}
\setlength{\labelsep}{-0.95em}
\itemsep 0.07cm
  \item[a)] We thoroughly investigate the implementation challenges of using RIS technology to achieve long-range, high-speed, and environmentally friendly underwater acoustic communications.
    \item[b)] An \textcolor{black}{synthetic reflection} scheme is developed for UA-RIS that allows to generation of reflected signals with arbitrary amplitude and phase shifts. This technique ensures precise steering of the reflected waves, directing them effectively while minimizing interference in unintended directions.
  \item[c)] Comprehensive validation of the UA-RIS was performed through tank and lake experiments \textcolor{black}{as well as COMSOL Multiphysics simulations}. The results demonstrate that the proposed system can significantly enhance signal reception quality in underwater environments, enabling long-range and high-speed acoustic communications.
\end{description}
\end{adjustwidth}
\vspace{0.1cm}

The structure of this paper is as follows: Section~\!\ref{sec:Related} reviews related work on UA-RIS. Section~\!\ref{sec:IQmod} examines the benefits of using \textcolor{black}{synthetic reflection} in UA-RIS. Section~\!\ref{sec:Hardware} provides a detailed description of the UA-RIS hardware design. Lastly, Section~\!\ref{sec:PerEva} presents an experimental validation of the system performance.

%%=========================================
%%=========================================

\section{Related Work}
\label{sec:Related}

\subsection{RIS in Terrestrial Radio Environments}
The advent of RF-RIS technology lies in the study of metamaterials, which are composed of periodic structures with scales that are smaller than the wavelength of interacting electromagnetic (EM) waves. These metamaterials can resonate with incident EM fields, consequently producing a specific EM response on the surface with a precise design of the surface architecture~\cite{munk2005frequency}.

Metasurfaces have the potential to minimize radar cross-sections (RCSs), thereby improving the stealthiness of aircraft and target objects. As studied in \cite{cui2014coding}, each unit on the metasurface can be configured to reflect incident EM waves in one of two states: in-phase or out-of-phase. By combining these states across the surface with different patterns, it is possible to minimize the intensity of the backscattered signal. 

Over the past decade, metasurfaces have experienced significant evolution, especially in their enhanced ability to manipulate the phase and amplitude of incident waves. Such advancements have led to the development of RIS, capable of forming arbitrary reflected beams in space. This innovation plays a pivotal role in improving communication quality and is regarded as a promising technology for the next generation of wireless networks~\cite{hassouna2023survey, naeem2023security}.

As introduced in \cite{boccia2002application}, the  phase response of a RIS can be electronically controlled by adding a varactor diode at the radiating edge of a patch antenna. This technique enables each RIS unit to achieve a tunable phase range of up to 180\textdegree. Expanding upon this idea, \cite{hum2007modeling} introduces a modification where microstrip patches are loaded with varactor diodes to enable broader beam forming. Such a modification changes the resonant frequency of each reflector unit, thereby affecting the phase of the scattered signal. The RIS units developed in their work exhibited more than 320\textdegree\ of phase agility at 5.5\,GHz, while sustaining a reflection efficiency above 70\%. The architecture of the RIS presented in \cite{hum2007modeling} has been applied in \cite{tan2016increasing} to enhance spectrum sharing at the 2.4\,GHz frequency. This enhancement is realized by improving the link quality between communicating parties while canceling out interference to other network participants through the RF-RIS.

The method involving the varactor to enable arbitrary phase shifts for extensive beamforming is further explored in \cite{pei2021ris}. Their study demonstrates that adjusting the varactor's bias voltage between 0\,V to 19\,V can produce a phase shift range from approximately $-$110\textdegree\ to 110\textdegree\ at 5.8\,GHz, while maintaining reflection efficiency above 70\%. In an indoor setting, the implemented RIS can overcome a 30\,cm thick concrete barrier and achieve a 26\,dB gain at the receiver. For outdoor applications, the system delivered a 14\,dB gain, facilitating high-speed communications across a distance of 500 meters.

In \cite{abeywickrama2020intelligent}, a practical phase shift model is proposed for RIS. Diverging from traditional models that typically assume the RIS can alter the phase of reflected waves between $-\pi$ and $\pi$, while maintaining a constant amplitude of 1, the new model takes into account the coupling between the amplitude and the phase shift of reflected waves, aligning with the experimental results presented in \cite{zhu2013active}. This refined model is advantageous for designing optimal reflection coefficients for RIS, as it can balance between the amplitude and the phase of signals reflected by each unit. This ensures that the aggregate of the signals reflected towards the receiver has the maximum SNR.

%=======================================
%=======================================
\subsection{RIS in Underwater Environments}
Inspired by achievements of metasurface in radio applications, the underwater acoustics sector is progressively embracing this new technology for its ability to manipulate the pattern of reflected waves. These artificial structures demonstrate proficiency in minimizing the scattering cross-section (SCS) of sonar systems. Such an advancement is of strategic importance, especially for enhancing stealth properties in military applications.

An example of the above application is the 2-bit coding metasurface developed in \cite{yu2021underwater}. This metasurface features four distinct types of elements. Each element is a steel plate punctured with four square holes at different depths, specifically engineered to shift the phase of reflected waves by 0, $\pi/2$, $\pi$, and $3\pi /2$. Through a meticulous arrangement of these elements, the design successfully attenuates the intensity of backscattered waves by at least 10\,dB, proving particularly effective within the 14.3\,kHz to 40.5\,kHz frequency range.

Beyond metamaterials, conventional piezoelectric ceramics can also electronically manipulate the phase and amplitude of reflected waves. For instance, the research conducted in \cite{eid2023enabling} showcases the development of an underwater backscatter array designed for long-distance passive acoustic communication. Utilizing the Van Atta array architecture, this system enables the reflected acoustic waves to automatically align with the direction of the incoming waves, eliminating the need for active components within the circuit. According to experimental data, the Van Atta backscatter array is capable of achieving a round-trip communication range exceeding 300 meters with a bit error rate (BER) of $10^{-3}$.

The design of a UA-RIS for high-data-rate and long-range communication is explored in \cite{sun2022high}. Each unit of the proposed UA-RIS comprises a piezoelectric coil sandwiched between two metal plates to reflect acoustic waves. The resonant impedance of the piezoelectric reflector is adjusted using a varactor and an MCU, allowing for continuous tuning of the acoustic unit's elasticity. This adjustment introduces additional phase shifts in reflected waves.

The study in \cite{wang2023designing} investigates wide-band beamforming using the UA-RIS architecture developed in \cite{sun2022high}. This approach aims to neutralize two types of dispersion effects: element dispersion within each piezoelectric reflector and array dispersion across the acoustic RIS. By doing so, the directivity of the reflected waves becomes frequency-independent, allowing wideband signals to achieve consistent beamforming results across various frequency components.The effectiveness of the proposed architecture is assessed using COMSOL Multiphysics and the Bellhop simulator. \textcolor{black}{While the proposed UA-RIS architecture demonstrates promising performance in simulation, its practical feasibility remains to be validated through real-world experiments.}

\textcolor{black}{The concept of RIS has also been extended to underwater optical communications to enable non-line-of-sight (NLOS) links when direct line-of-sight (LOS) paths are obstructed by obstacles such as schools of fish, submarines, or seamounts \cite{odeyemi2020performance, ramavath2024performance}. In \cite{odeyemi2020performance}, an analytical study of an RIS-assisted underwater optical communication system is presented. It models a decode-and-forward RIS relay under mixed Exponential-Gamma oceanic turbulence and Rayleigh fading, deriving closed-form expressions for end-to-end outage probability and average BER. More recently, the research in \cite{ramavath2024performance} tackled the challenge of  ``skip-zones'' at the air-water interface by placing RIS at the boundary of an free-space optical communication link. By incorporating models of Gamma-Gamma atmospheric turbulence, generalized Gamma oceanic turbulence, and pointing errors, it demonstrates that RIS can bridge up to 30 meters of optical dead zone and improve BER by five orders of magnitude at 1\,Gbps.}

\textcolor{black}{Parallel-relay designs have also been explored in \cite{yadav2023performance}. This research proposed a dual-hop system with separate free-space optical communication and underwater wireless optical communication decode-and-forward relays at different depths, selecting the best branch by instantaneous SNR. Their analysis under worst-case turbulence and misalignment predicts a $\sim$3\,dB gain over single-relay schemes at $10^{-4}$ outage, with free-space optical communication dominating in clear-water conditions. On the hardware front, \cite{xu2024fully} demonstrated a 16$\times$16 liquid-crystal spatial light modulator RIS with integrated aperture-averaging lens and phased-array beamformer. In lab tests at 10\,Mbps over 15\,m, adaptive beam steering achieved BER less than $10^{-5}$ and a fourfold SNR improvement compared to no-RIS.}

\textcolor{black}{Although underwater optical communication offers high data rates, its effective range is typically under 100 meters~\cite{odeyemi2020performance, ramavath2024performance, yadav2023performance, xu2024fully}. In contrast, acoustic RIS leverages lower frequencies and longer wavelengths to achieve communication to kilometer scales even in turbid, particle-rich waters that rapidly attenuate optical beams, making UA-RIS ideal for low-power, long-range underwater communications in challenging environments.}

\textcolor{black}{Our work presents two key innovations that distinguish it from prior efforts on UA-RIS~\cite{wang2023designing, sun2022high}. First, we propose a novel approach to phase control in UA-RIS that addresses practical limitations associated with existing architectures. While previous studies have explored varactor-based impedance tuning, an approach well-suited to RF applications but challenging at acoustic frequencies due to the significantly larger capacitance variation required. To address this challenge, we introduce a synthetic reflection technique that achieves phase control by combining orthogonal components generated by two independently controlled reflectors. This design allows for flexible and accurate phase adjustment using readily available components. Second, whereas earlier works have primarily focused on theoretical analysis and simulation, our study includes comprehensive experimental validation of the proposed architecture. We conduct both tank experiments and open-water lake tests to evaluate the effectiveness and feasibility of the proposed architecture in real-world conditions. To the best of our knowledge, this is the first field-deployed and experimentally verified UA-RIS, providing a critical step toward practical underwater acoustic communication using intelligent surfaces.}

%%=========================================
%%=========================================

\section{\textcolor{black}{Synthetic Reflection} for UA-RIS}
\label{sec:IQmod}
This section explores the \textcolor{black}{synthetic reflection} technique utilized in our UA-RIS. We begin by examining the challenge of implementing arbitrary phase shifts within UA-RIS, followed by a description of how \textcolor{black}{synthetic reflection} addresses this issue. Subsequently, we discuss the benefits of \textcolor{black}{synthetic reflection}, highlighting its feasibility, flexibility, and interference reduction.

%=================================================
\subsection {Challenges of Implementing Arbitrary Phase-Shift in UA-RIS}
\label{sec:MotiIQ}
In an ideal RIS, each reflecting unit should independently and passively control both the amplitude and phase of reflected waves. However, implementing this functionality presents challenges in practical systems due to hardware constraints. As a result, most existing RF-RIS designs maintain a fixed amplitude and focus on modifying the phase of the reflected wave. This is typically achieved by adjusting the load impedance of each reflection unit using varactor diodes \cite{pei2021ris, abeywickrama2020intelligent}. Unfortunately, this approach is not viable for UA-RIS due to the low frequency of acoustic signal, as will be analyzed next.

When a wave propagates through mediums with discontinuous impedances, reflection occurs. The reflection coefficient, denoted by $\Gamma$, is calculated using:
\begin{equation}
\label{eq:0aw2}
  \Gamma = \displaystyle\frac{Z_L-Z_0}{Z_L+Z_0},
\end{equation}
where $Z_L$ represents the load impedance, and $Z_0$  is the characteristic impedance of the transmission line.

According to (\ref{eq:0aw2}), an RIS with a characteristic impedance of 50\,$\Omega$, typical for RF transmission lines, requires load impedances of $-121j\,\Omega$ and $-50j\,\Omega$ to generate reflected waves with the same amplitude as the incident wave but with phase shifts of $-45^\circ$ and $-90^\circ$, respectively. Assuming a signal frequency of 30\,GHz is selected for mmWave communication, these load impedances correspond to capacitances of approximately 44\,fF and 106\,fF, respectively. Commercial varactor diodes, such as the ZC836A~\cite{830ser2007zetex} and NTE618~\cite{nte2012nte}, can continuously vary capacitance within this range by adjusting the reverse bias voltage applied to their p-n junctions, thereby enabling arbitrary phase shifts of the reflected waves.

However, the frequencies used for middle- and long-range applications typically employs acoustic frequencies below 50\,kHz. Under these conditions, the capacitance of a varactor diode must be varied substantially to effect a desired phase shift in the reflected wave. \textcolor{black}{For example, under the same 50\,$\Omega$ impedance-matching conditions used previously, reducing the source frequency to 50\,kHz (for mid-range acoustic communication) requires the RIS load capacitance to be tuned from 26\,nF to 64\,nF in order to shift the reflection phase from $-45^\circ$ to $-90^\circ$. If the frequency is further reduced to 25\,kHz (for long-range communication), the capacitance must be adjusted from 52\,nF to 128\,nF to obtain the same phase-shift range. Such capacitance variation far exceeds the tuning capability of commercial varactor diodes, which generally have capacitance values below 1\,nF, and thus a new method is required to realize arbitrary phase shifts in UA-RIS.}

%=================================================
\subsection {Implementation of \textcolor{black}{Synthetic Reflection}}
\label{sec:ImpIQ}
In the proposed UA-RIS, we employ \textcolor{black}{synthetic reflection} scheme to generate reflected waves with arbitrary amplitude and phase shift. To implement this scheme, the MCU automatically pairs acoustic reflectors according to the angles of reflected waves. Subsequently, one reflector in each pair produces an in-phase or antiphase component, shifting the phase of the reflected waves by $0$\textdegree\ or $180$\textdegree. The other reflector generates a quadrature component, shifting the phase of the reflected waves by $90$\textdegree\ relative to the incident signal. By precisely controlling the amplitude of the in-phase, antiphase, and quadrature components, the UA-RIS can synthesize reflected waves with any desired phase in space, as dictated by trigonometric identities:
\begin{equation}\label{eq:js23}
    \begin{array}{lll}
       \sin(\omega t+\phi)&\!\!\!=\!\!\!&\sin\phi\cos\omega t + \cos\phi\sin\omega t\\
       &\!\!\!\triangleq\!\!\!& A_I \cos\omega t + A_Q\sin\omega t,
    \end{array}
\end{equation}
where $\omega$ is angular frequency; $t$ is time; $\phi$ is arbitrary phase shift between $-\frac{\pi}{2}$ and $\frac{\pi}{2}$; $A_I$ is the amplitude of in-phase ($A_I\!\geq\!0$) or antiphase component($A_I\!<\!0$); $A_Q$ is the amplitude of the quadrature component.

\begin{figure}[htb]
\centerline{\includegraphics[width=6.0cm]{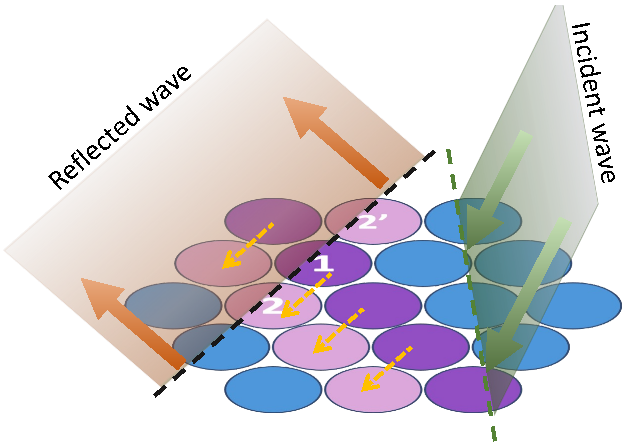}}
  \caption{\textcolor{black}{synthetic reflection} for arbitrary phase shift.}\label{fig:IQmod}
\end{figure}

Fig.~\!\ref{fig:IQmod} illustrates how \textcolor{black}{synthetic reflection} technology operates. In the diagram, each circle represents an acoustic reflector, the size of which is one wavelength to efficiently reflect the incident waves. In the figure, the green plane indicates the direction of the incident wave, while the orange plane designates the target direction for the reflected waves.

To execute \textcolor{black}{synthetic reflection}, only reflectors located on the wavefront of the reflected wave (as marked by the black dotted line in the figure) can be paired. For example, reflector 1 can be paired with reflector 2 or 2$^\prime$. With this configuration, the projection positions of the paired reflectors along the direction of the reflected waves align. Consequently, positional differences between paired reflectors do not result in additional phase shifts in the direction of the reflected wave. This setup enables the paired reflectors to function collectively as a single reflector, facilitating precise beam steering.

It is important to note that the phase of the incident wave received by each reflector in a pair may differ. This discrepancy arises because the projection positions of the paired reflectors along the wavefront of the incident signals do not always align, as depicted in Fig.~\!\ref{fig:IQmod}, where the green dotted line (representing the wavefront of the incident wave) is not parallel to the black dotted line. Therefore, the UA-RIS must correct these phase differences to ensure that the reflected waves generated by the pair of reflectors remain orthogonal, which is crucial for effective \textcolor{black}{synthetic reflection}.

To address the aforementioned issue, the UA-RIS must tailor the reflection coefficients of each reflector based on the angle of the incident wave. Taking the pair of reflectors 1 and 2 in Fig.~\!\ref{fig:IQmod} as an example, assume the incident waves received by reflector 1 and reflector 2 are $\cos(\omega t+\phi_1)$ and $\cos(\omega t+\phi_2)$, respectively. The MCU then adjusts the impedance of the load network to set the reflection coefficient of reflectors 1 and 2 to $A_1$ and $jA_2$, respectively, where $j$ represents the imaginary component with $90^\circ$ phase shift. This configuration results in the sum of the acoustic waves reflected by reflectors 1 and 2 being:
\begin{equation}\label{eq:8a2h}
    \begin{aligned}
       & A_1\cos(\omega t+\phi_1) \!+\! A_2\cos(\omega t+\frac{\pi}{2}+\phi_2) \\
       =& \;\;\;\left(A_1\cos\phi_1 - A_2\sin\phi_2\right)\cos\omega t \\
       		&-\!\left(A_1\sin\phi_1 + A_2\cos\phi_2\right)\sin\omega t.
    \end{aligned}
\end{equation}
Comparing (\ref{eq:8a2h}) with (\ref{eq:js23}) reveals that $\left(A_1\cos\phi_1-A_2\sin\phi_2\right)$ is the amplitude of the in-phase or antiphase component, while $-\left(A_1\sin\phi_1 + A_2\cos\phi_2\right)$ is the amplitude of the quadrature components for \textcolor{black}{synthetic reflection}.

To direct the reflected signal towards the desired direction, assume that the waveform synthesized by reflectors 1 and 2 is represented as:
\begin{equation}
\label{eq:nc3t}
  \cos(\omega t+\phi_r) = \cos\phi_r\cos\omega t - \sin\phi_r\sin\omega t,
\end{equation}
where $\phi_r \!\in\! [-\frac{\pi}{2}, \frac{\pi}{2}]$. Then, by combining (\ref{eq:8a2h}) and (\ref{eq:nc3t}), the coefficients $A_1$ and $A_2$ of the reflectors 1 and 2 can be derived as follows:
\begin{equation}
\label{eq:kas4}
A_1=\displaystyle\frac{\cos(\phi_r-\phi_2)}{\cos(\phi_1-\phi_2)} \qquad  \text{and} \qquad 
A_2=\displaystyle\frac{\sin(\phi_r-\phi_1)}{\cos(\phi_1-\phi_2)}.
\end{equation}
Using (\ref{eq:kas4}), the MCU can calculate the reflection coefficient for each reflector based on the angles of the incident and reflected waves. This enables the synthesis of reflected waves with arbitrary phases, facilitating precise beam steering as required.

%%=========================================
%%=========================================

\section{Hardware Architecture}
\label{sec:Hardware}
This section outlines the hardware design of UA-RIS, beginning with an introduction to the structure of the acoustic reflector and the circuitry designed for efficient \textcolor{black}{synthetic reflection}. Following this, we will discuss several critical considerations essential for real-world implementation.

%=================================================
\subsection {Structure and Circuit of Acoustic Reflector}
\label{sec:StruAco}

\subsubsection{Reflector Structure}
Our work concentrates on acoustic signals within the mid-frequency range between 10\,kHz and 50\,kHz. As a result, the Tonpilz structure is selected for constructing the reflection unit, due to its superior electroacoustic conversion efficiency and effective impedance matching between the piezoelectric materials and water. This configuration allows for the flexible adjustment of the amplitude and phase of reflected waves.

\begin{figure}[htb]
\centerline{\includegraphics[width=6.5cm]{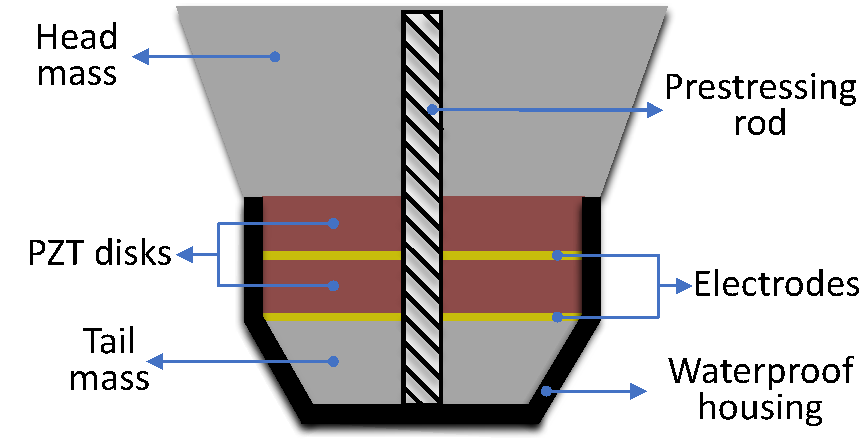}}
  \caption{Tonpilz reflector used in our UA-RIS.}\label{fig:Reflector}
\end{figure}

Fig.~\!\ref{fig:Reflector} depicts the structure of the reflector used in our UA-RIS, consisting of a head mass, two PZT disks equipped with electrodes, a prestressing rod, and a tail mass. The head mass, a round cone cast from aluminum alloy, ensures effective oscillation along the central axis without placing undue stress on any component of the reflector. Situated behind the head mass are two PZT disks, strung together with a prestressing rod and separated by copper sheets that link to the subsequent circuit to regulate the reflector's load impedance. The tail mass at the rear reflects backward-traveling vibrations, ensuring that the acoustic wave is projected forward.

In practical applications, achieving sufficient directional gain for a high-performance UA-RIS typically necessitates a substantial array of acoustic reflectors, impacting the size and cost of the system. Consequently, our UA-RIS is equipped with only two PZT plates per reflector, each with 38\,mm in diameter and 5\,mm in height. This build significantly reduces size, wight, and cost, facilitating widespread deployment of UA-RIS in marine environments.

\vspace{0.2cm}
\subsubsection{Circuit Design}
\begin{figure}[htb]
\centerline{\includegraphics[width=9cm]{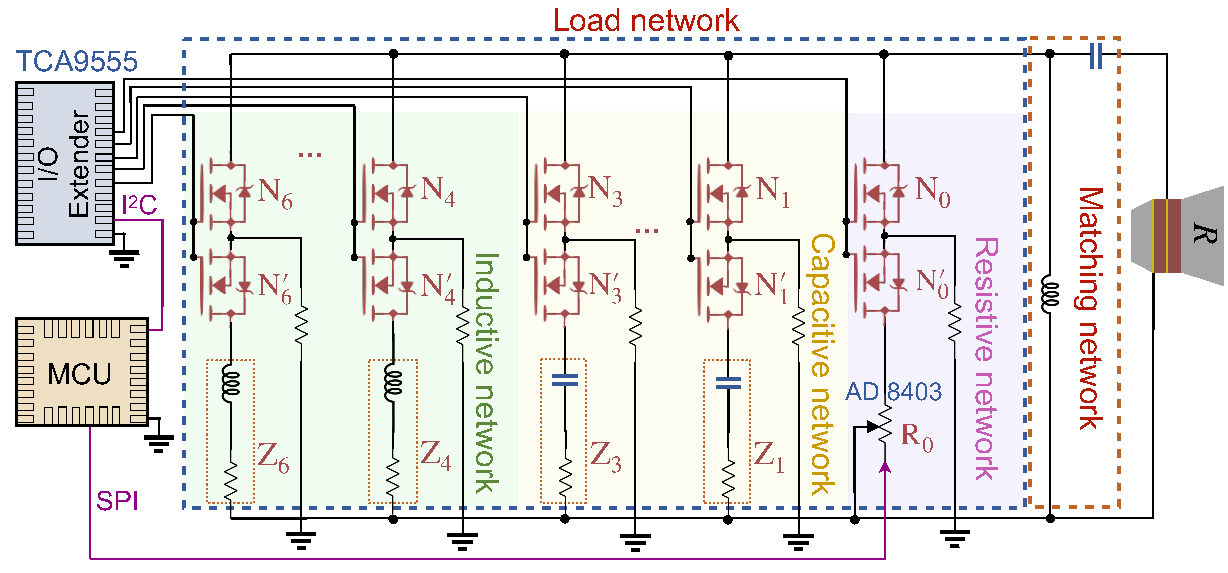}}
  \caption{Architecture of the UA-RIS.}\label{fig:Circuit}
\end{figure}

Fig.~\!\ref{fig:Circuit} presents the circuit architecture of the proposed UA-RIS, consisting three primary components: a matching network, a load network, and a control \& processing unit. Next, we will delve into a detailed explanation of each component.

In the circuit, the matching network aligns the impedance of all reflectors to a real number, which is represented as $Z_0$ in (\ref{eq:0aw2}). Subsequently, the reflection coefficient for each reflector is modulated by linking different stages of the load network to the matching circuit. The connection of each stage is controlled through a pair of n-type metal-oxide-semiconductor (NMOS) transistors, whose ON/OFF states are regulated by TCA9555 I/O extenders and an ATmega256RFR2 MCU.

The load network comprises three subnetworks: a resistor network, a capacitive network, and an inductive network. The resistor network has only one stage designed to generate the in-phase or antiphase component required for \textcolor{black}{synthetic reflection}. Conversely, both the capacitive and inductive networks feature three stages each, dedicated to producing the quadrature component for \textcolor{black}{synthetic reflection}.

Resistor network contains an AD8403AR50 programmable potentiometer, whose resistance is managed by the MCU via the serial peripheral interface (SPI) protocol. Let $R_0$ represent the resistance of this potentiometer. According to the reflection coefficient presented in (\ref{eq:0aw2}), when the load connected to the matching circuit is non-reactive and $Z_0$ is a real number, $\Gamma$ has no imaginary component. In this circumstance, the magnitude of the reflection coefficient varies between 0 and 1, while the phase shift of the reflected wave remains fixed at 0\textdegree\ for $R_0 \geq Z_0$, or 180\textdegree\ for $R_0 < Z_0$. \textcolor{black}{For instance, if the characteristic impedance of the transmission line is $Z_L\!=\!50\,\Omega$, then according to (\ref{eq:0aw2}), the reflection coefficient $\Gamma$ is 0.33 when $R_0\!=\!100\,\Omega$ and $-$0.33 when $R_0\!=\!25\,\Omega$. In both cases, the reflected signals have identical amplitudes but are 180\textdegree\ out of phase.}

According to (\ref{eq:0aw2}), when $Z_0$ is a real number, the load impedance must be a complex number to generate the quadrature component necessary for \textcolor{black}{synthetic reflection}. In this scenario, it is impossible to continuously adjust the magnitude of the reflection coefficient by simply altering the resistance of the load impedance. To tackle this issue, in our design, the capacitive network includes three stages, designated as $Z_1$ to $Z_3$ in Fig.~\!\ref{fig:Circuit}. Connecting different stages of this network to the matching circuit yields a reflection coefficient with a consistent phase of $-$90\textdegree\ and three selectable amplitudes increasing stepwise from 0.3 to 0.9. Conversely, $Z_4$ to $Z_6$ form a three-stage inductive network that produces a reflection coefficient with a consistent phase of 90\textdegree, and three optional amplitudes ranging from 0.3 to 0.9.

\textcolor{black}{For example, consider an acoustic signal with a frequency of 28\,kHz and a characteristic impedance $Z_0\!=\!50\,\Omega$. Suppose a capacitive subnetwork consists of a resistance of 23.5\,$\Omega$ and a capacitance of 128.6\,nF. The resulting load impedance is calculated as
\begin{equation}
\label{eq:js8n}
	\begin{array}{lll}
  		Z_L \!\!\!\!&=&\!\!\! 23.53 + \displaystyle\frac{1}{2\pi j\!\times\!28\,\text{kHz}\!\times\!128.6\,\text{nF}}\\
  		\!\!\!\!&=&\!\!\!23.53-44.12j\Omega.
  \end{array}
\end{equation}
Substituting (\ref{eq:js8n}) into (\ref{eq:0aw2}), the reflection coefficient is given by
\begin{equation}
\label{eq:nby6}
  \Gamma = \displaystyle\frac{Z_L-Z_0}{Z_L+Z_0} =\displaystyle\frac{-26.47-44.12j}{73.53-44.12j}\approx-0.6j.
\end{equation}
In this case, the reflected signal experiences a phase shift of $-$90\textdegree\ and an amplitude that is 60\% of the incident signal.}

With the architecture depicted in Fig.~\!\ref{fig:Circuit}, the MCU can effectively manipulate both the phase and magnitude of the reflection coefficient by adjusting the resistance of the potentiometer and the ON/OFF status of the NMOS transistors in each stage of the load network. In the following section, we will introduce the selection of $Z_0$ and discuss the strategy for managing the high dynamic range of the acoustic signal in the circuit design.

%=================================================
\subsection {High-Impedance Matching}
\label{sec:HigImp}
In the design of the load network for UA-RIS, we match the reflector's impedance to 1\,k$\Omega$ (i.e., $Z_0\!=\!1$\,k$\Omega$), a value substantially higher than the standard $50\,\Omega$ or $75\,\Omega$ commonly employed in radio communication systems. This significant divergence is attributed to the inherent resistance introduced by the wiper of the programmable potentiometer.

In an ideal potentiometer, the resistance between a terminal and the wiper should vary continuously from 0 to $R_{max}$, where $R_{max}$ is the resistance between the two terminals. However, in a real programmable potentiometer, an array of MOSFETs is usually employed to emulate the function of a mechanical wiper for resistance adjustment. These MOSFETs inherently exhibit a certain level of resistance.

In order to precisely control the phase and amplitude of the quadrature component for accurate \textcolor{black}{synthetic reflection}, the wiper resistance of programmable potentiometers cannot be ignored. For instance, models such as the AD8403~\cite{analog2010channel} and AD5245~\cite{analog2022position} typically exhibit a wiper resistance of about 50\,$\Omega$, while the MAX5427 has a resistance of 100\,$\Omega$~\cite{maxim2005one}. Consequently, the minimum resistance of the load network in (\ref{eq:0aw2}) does not approach zero but remains significantly higher. In scenarios where the impedance of the reflector is matched to 50\,$\Omega$ (i.e., $Z_0\!=\!50$\,$\Omega$), the reflection coefficient consistently remains positive. As a result, the reflector will be incapable of generating an antiphase wave.

To address the aforementioned problem, we match the impedance of the acoustic reflector to 1\,k$\Omega$, significantly higher than the inherent wiper resistance of the programmable potentiometer. This configuration allows the MCU to flexibly adjust the amplitude of the quadrature component between $-$1 and 1 by programming the resistance of the potentiometer.

%=================================================
\subsection {Highly Dynamic Incident Waves}
\label{sec:HigDyn}
In practical scenarios, the distance between the UA-RIS and the acoustic transmitter may be unknown or can change over time in mobile applications. As a result, the intensity of the acoustic waves reaching the UA-RIS can vary significantly. To ensure the reliability of the UA-RIS, it is crucial that the hardware performs consistently across environments with both high and low energy densities.

Acoustic signals naturally consist of periodic compressions and rarefactions within the medium. Consequently, the output from a reflector is always alternating current (AC). Therefore, the electronic switches in a UA-RIS must be capable of handling bidirectional current flow. Taking NMOS transistors as an example, when the incident power is weak, a single MOSFET can be utilized as an electronic switch to control the ON and OFF states of the load. However, as the intensity of the incident power increases, the intrinsic body diode of the NMOS transistor conducts during the negative half-cycle of the signal. Under these conditions, the load remains connected for half the signal cycle even if the gate-to-source voltage of the transistor falls below a threshold.

To mitigate the aforementioned issue, the we pairs MOSFETs as an electronic switch. As depicted in Fig.~\!\ref{fig:Circuit}, the source terminal of the upper NMOS transistor is connected to the drain terminal of the lower one. This arrangement allows the body diode of the lower transistor to effectively prevent any reverse current in the load network. However, this setup necessitates at least an additional transistor and a pull-down resistor, denoted by $R_{pd}$ in the figure, for each stage of the load network. Consequently, this escalates the overall cost of the UA-RIS, especially when a large number of acoustic reflectors are incorporated to support long-distance underwater communication.

%%=========================================
%%=========================================

\section{Performance Evaluation}
\label{sec:PerEva}
This section evaluates the performance of the proposed UA-RIS through experiments. Initially, we examine the power consumption of the system. Subsequently, the results collected from \textcolor{black}{COMSOL simulation, as well as controlled tank tests and open-water lake experiments} are presented to demonstrate the viability of the UA-RIS in real-world applications.

%=================================================

\subsection {Power Consumption}
\label{sec:PowCon}
Although the UA-RIS does not transmit acoustic signals, it still requires an energy supply to maintain the operation of active components for \textcolor{black}{synthetic reflection}. In our implementation, the power consumption of the system primarily originates from four components: the MCU, twelve I/O extenders, and six programmable potentiometers. In this section, we thoroughly analyze the power consumption of UA-RIS operating in various states.

In the absence of target waves, the UA-RIS remains in standby mode to save the energy. In this mode, the current consumption for the ATmega256RFR2 MCU, TCA9555 I/O extender with high inputs, and AD8403AR50 potentiometer at a 2\,V supply voltage are 650\,nA, 500\,nA, and 5\,$\mu$A, respectively. Consequently, the total power consumption of the system is only 73.3\,$\mu$W. However, upon detection of acoustic signals, UA-RIS transitions to an active mode. In this scenario, the power consumption of the system is divided into two stages: (a) the impedance adjustment phase, and (b) the impedance maintenance phase.

In the impedance adjustment phase, the MCU controls the I/O extender and AD8403AR50 potentiometer to finely tune the load impedance in alignment with the desired direction of the reflected waves. In the worst case, all load impedances shift from reactance to resistance, necessitating concurrent programming of all potentiometers. To achieve this, the MCU sends three bytes of data (chip address + register address + output value) via the $\text{I}^2\text{C}$ protocol to manage the I/O extender for each channel. Additionally, it transmits two bytes (channel ID and value) through the SPI protocol to set the resistance for each potentiometer. As a result, the MCU issues a total number of 72 bytes and 48 bytes through the $\text{I}^2\text{C}$ and SPI protocols, respectively.

After completing the load configuration, the circuit transitions into the impedance maintenance phase. In this phase, the MCU can enter a sleep mode, while the I/O extender and potentiometer continue to operate, maintaining the states of electronic switches and load impedance during a signal period for beam steering. At this time, the power consumption of the UA-RIS lies between that of the standby mode and the first phase of the active mode.

\begin{figure}[htb]
\centerline{\includegraphics[width=7.0cm]{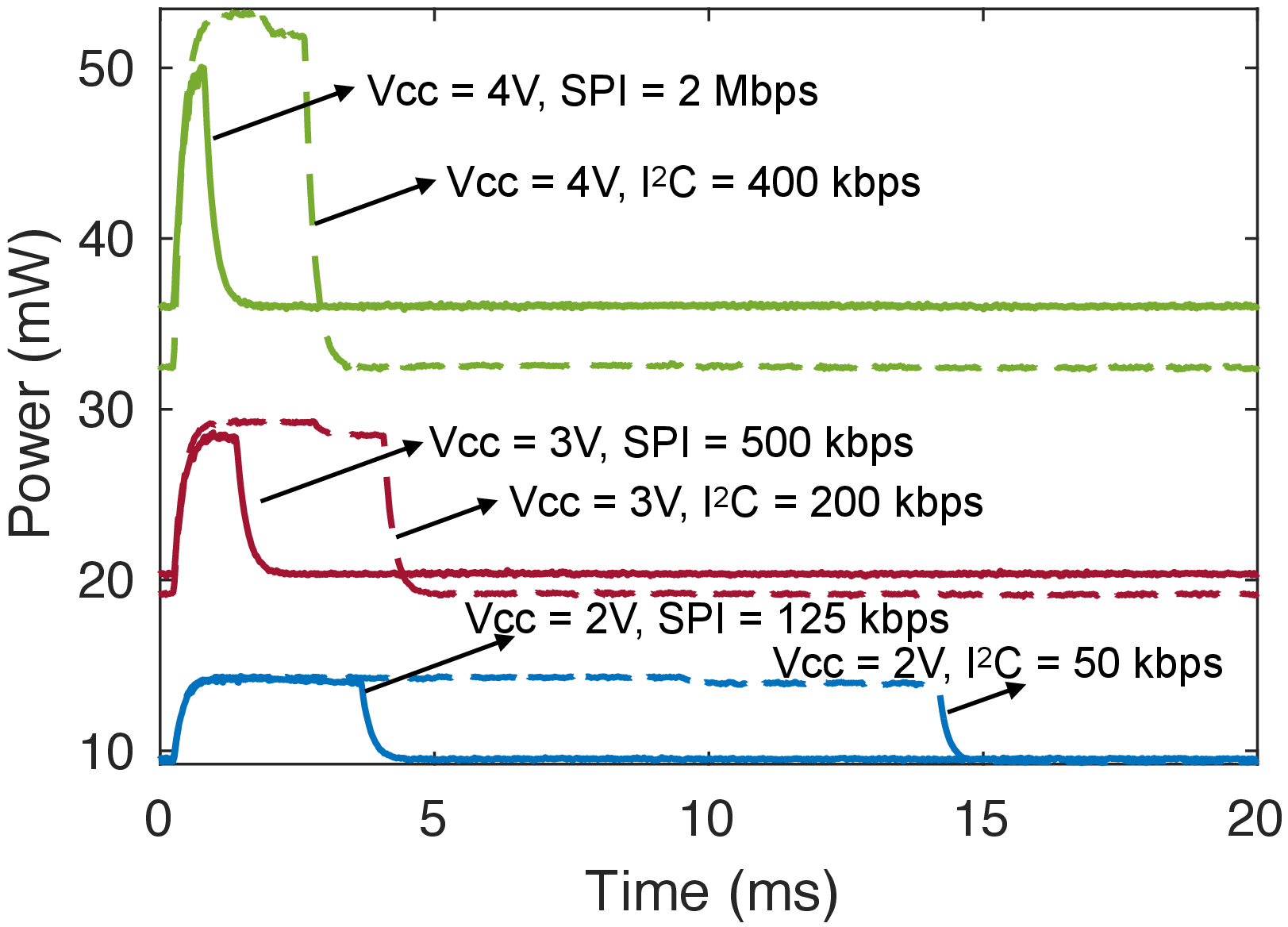}}
  \caption{Power consumption of the UA-RIS at various supply voltages.}\label{fig:pwrWaveform}
\end{figure}

Fig.~\!\ref{fig:pwrWaveform} displays the power consumption of the UA-RIS in the active mode, in which the MCU updates the outputs of the I/O extender and adjusts the resistance of the potentiometer across all 24 reflectors. For this experiment, the MCU operates at 8\,MHz, and the system is powered by a Keysight EDU36311A DC supply. In the figure, the peaks of the waveform corresponding to $\text{I}^2\text{C}$ or SPI communications occur at a specified supply voltage during the impedance adjustment phase, while the subsequent data illustrate the power consumption during the impedance maintenance phase.

As depicted in Fig.~\!\ref{fig:pwrWaveform}, the overall system power consumption in active mode varies significantly with changes in the supply voltage, denoted by $V_{cc}$. For instance, at a $V_{cc}$ of 2\,V, the peak power consumed for $\text{I}^2\text{C}$ or SPI serial communication is only 14.2\,mW. However, when $V_{cc}$ is increased to 4\,V, the power consumption rises to 53\,mW, a 3.7-fold increase compared to the lower supply voltage. Similarly, power consumption during the impedance maintenance phase is also proportional to $V_{cc}$. At 2\,V, maintaining the load impedance of 24 channels consumes only 9.3\,mW, whereas at 4\,V, this consumption escalates to 35.3\,mW.

\begin{table}[htp]
\footnotesize
\centering
\setlength{\tabcolsep}{4pt} % reduced horizontal padding
\caption{\MakeUppercase{Energy/power consumption in the active mode}}
\label{tab:EnyCon}
\begin{tabular}{|
>{\columncolor[HTML]{EFEFEF}}c|
>{\columncolor[HTML]{D0D0D0}}c
>{\columncolor[HTML]{D0D0D0}}c
>{\columncolor[HTML]{D0D0D0}}c
>{\columncolor[HTML]{B4B4B4}}c
>{\columncolor[HTML]{B4B4B4}}c
>{\columncolor[HTML]{B4B4B4}}c|
>{\columncolor[HTML]{EAEAEA}}c|}
\hline
\multicolumn{1}{|l|}{\cellcolor[HTML]{FFFFFF}} &
\multicolumn{6}{c|}{\cellcolor[HTML]{EEEEEE}Phase I} &
\cellcolor[HTML]{EAEAEA} \\ \cline{1-7}
\cellcolor[HTML]{EFEFEF} &
\multicolumn{3}{c|}{\cellcolor[HTML]{E1E1E1}$\text{I}^2\text{C}$ ($\mu$J)} &
\multicolumn{3}{c|}{\cellcolor[HTML]{E1E1E1}SPI ($\mu$J)} &
\cellcolor[HTML]{EAEAEA} \\ \cline{2-7}
\multirow{-2}{*}{\cellcolor[HTML]{EFEFEF}Vcc} &
\multicolumn{1}{c|}{\cellcolor[HTML]{D0D0D0}\begin{tabular}[c]{@{}c@{}}50\\(kbps)\end{tabular}} &
\multicolumn{1}{c|}{\cellcolor[HTML]{D0D0D0}\begin{tabular}[c]{@{}c@{}}200\\(kbps)\end{tabular}} &
\multicolumn{1}{c|}{\cellcolor[HTML]{D0D0D0}\begin{tabular}[c]{@{}c@{}}400\\(kbps)\end{tabular}} &
\multicolumn{1}{c|}{\cellcolor[HTML]{B4B4B4}\begin{tabular}[c]{@{}c@{}}125\\(kbps)\end{tabular}} &
\multicolumn{1}{c|}{\cellcolor[HTML]{B4B4B4}\begin{tabular}[c]{@{}c@{}}500\\(kbps)\end{tabular}} &
\begin{tabular}[c]{@{}c@{}}2\\(Mbps)\end{tabular} &
\multirow{-3}{*}{\cellcolor[HTML]{EAEAEA}\begin{tabular}[c]{@{}c@{}}Phase II\\(mJ)\end{tabular}} \\ \hline
2 V &
\multicolumn{1}{c|}{\cellcolor[HTML]{D0D0D0}198.5} &
\multicolumn{1}{c|}{\cellcolor[HTML]{D0D0D0}56.6} &
\multicolumn{1}{c|}{\cellcolor[HTML]{D0D0D0}38.4} &
\multicolumn{1}{c|}{\cellcolor[HTML]{B4B4B4}48.9} &
\multicolumn{1}{c|}{\cellcolor[HTML]{B4B4B4}15.3} &
8.8 &
9.3 \\ \hline
3 V &
\multicolumn{1}{c|}{\cellcolor[HTML]{D0D0D0}397.0} &
\multicolumn{1}{c|}{\cellcolor[HTML]{D0D0D0}112.4} &
\multicolumn{1}{c|}{\cellcolor[HTML]{D0D0D0}73.7} &
\multicolumn{1}{c|}{\cellcolor[HTML]{B4B4B4}98.1} &
\multicolumn{1}{c|}{\cellcolor[HTML]{B4B4B4}33.6} &
18.0 &
19.9 \\ \hline
4 V &
\multicolumn{1}{c|}{\cellcolor[HTML]{D0D0D0}694.7} &
\multicolumn{1}{c|}{\cellcolor[HTML]{D0D0D0}198.4} &
\multicolumn{1}{c|}{\cellcolor[HTML]{D0D0D0}127.7} &
\multicolumn{1}{c|}{\cellcolor[HTML]{B4B4B4}172.5} &
\multicolumn{1}{c|}{\cellcolor[HTML]{B4B4B4}58.6} &
31.8 &
35.3 \\ \hline
\end{tabular}
\end{table}

Table~\ref{tab:EnyCon} details the energy spent in the two phases of active mode, where the length of the acoustic signal is 1 second. The results demonstrate that higher data rates can substantially reduce transmission time, thus lowering energy consumption in phase 1. For example, at a 2\,V supply voltage, with $\text{I}^2\text{C}$ and SPI baud rates at 50\,kbps and 125\,kbps, respectively, the energy consumed for serial communication are 198.5\,$\mu$J and 48.9\,$\mu$J, respectively, totaling 247.4\,$\mu$J for the first phase. In contrast, increasing the baud rate of $\text{I}^2\text{C}$ to 400\,kbps, and SPI to 2\,Mbps, the maximum speeds supported by the ATmega256RFR2 MCU, reduces the energy consumption in phase 1 to 47.2\,$\mu$J (38.4\,$\mu$J + 8.8\,$\mu$J), which is only one-fifth of the earlier scenario.

Due to the long preambles associated with slow data rates, the duration of underwater acoustic signals is significantly longer than that of RF signals~\cite{luo2014challenges}. Consequently, the UA-RIS must maintain the load impedance for hundreds of milliseconds or longer to accommodate a complete signal period, resulting in substantial energy consumption. According to Table~\ref{tab:EnyCon}, when the supply voltage is 2\,V and serial communication occurs at maximum speed, over 95\% of the energy is consumed in phase 2. Increasing the supply voltage to 4\,V raises this proportion to 99.6\%.

%=================================================

\begin{figure}[htb]
\centerline{\includegraphics[width=6.5cm]{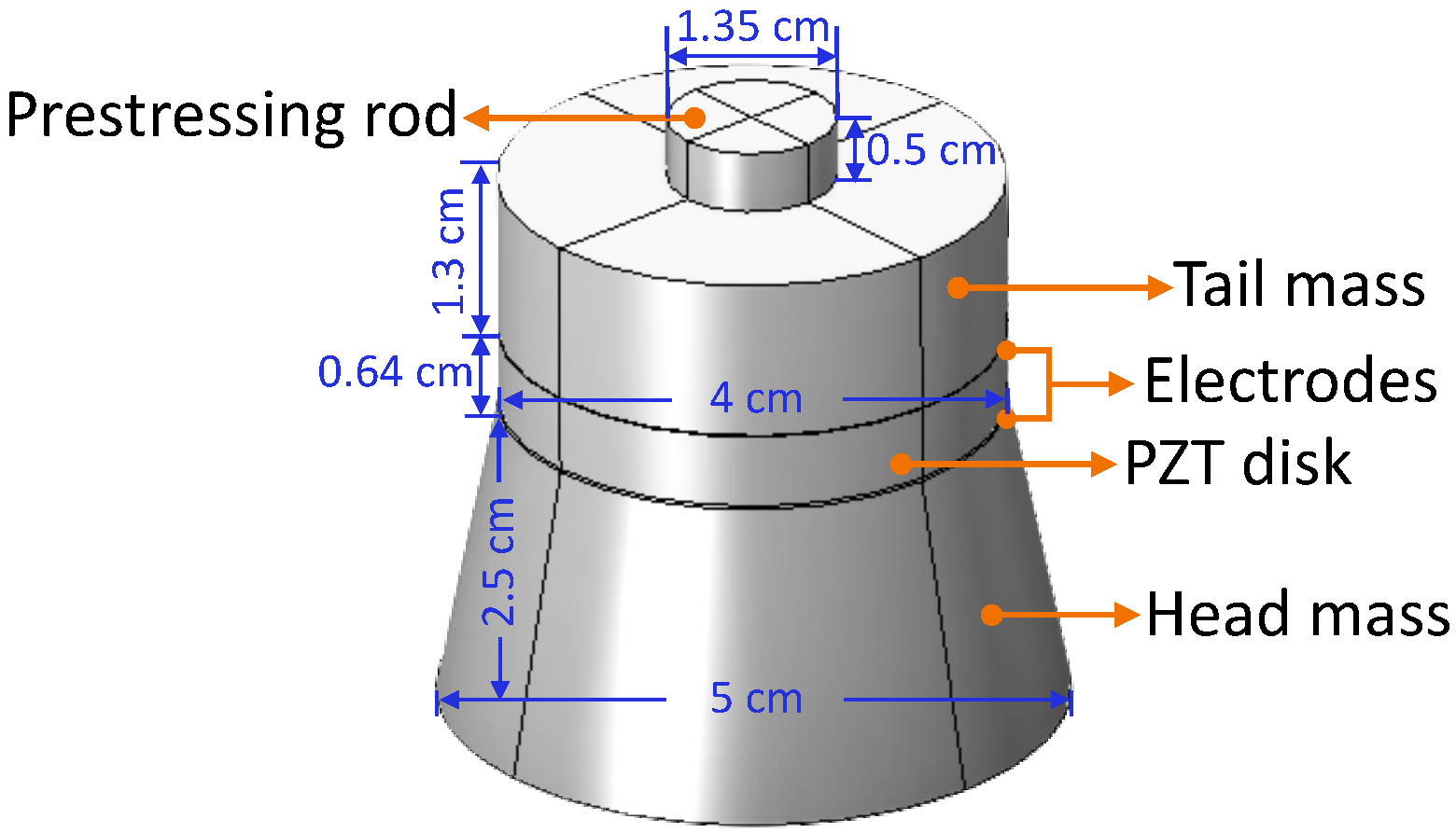}}
  \caption{3D structure of the acoustic reflector constructed for the simulation.}\label{fig:3DStr}
\end{figure}

\textcolor{black}{
\subsection {COMSOL Simulation Results}
\label{sec:SimRes}
To assess the performance of the UA-RIS, simulations are carried out using COMSOL Multiphysics. The 3D structure of the acoustic reflector is illustrated in Fig.~\!\ref{fig:3DStr}. The head mass is constructed from aluminum, while the tail mass and prestressing rod are made of AISI 4340 steel. PZT-4 is employed as the acoustic vibrational element.}

\textcolor{black}{In the simulation, a monochromatic plane wave at 28\,kHz is used as the signal source. To enhance simulation accuracy, a five-step procedure is followed. First, the impedance of  the PZT disk is determined by measuring the ratio of open-circuit voltage to short-circuit current. Based on this impedance value, appropriate matching circuit is constructed. A load is then introduced, and the resulting voltage response under the plane wave excitation is recorded. This measured voltage, together with the reflection coefficients of the load, is used to compute the reflected voltage. Finally, with the incident wave deactivated, the computed reflected voltage is applied to the matching circuits, and the resulting pressure waveform is recorded to represent the signal reflected by the structure under the specified load conditions.}

\begin{figure}[htb]
\centerline{\includegraphics[width=8.3cm]{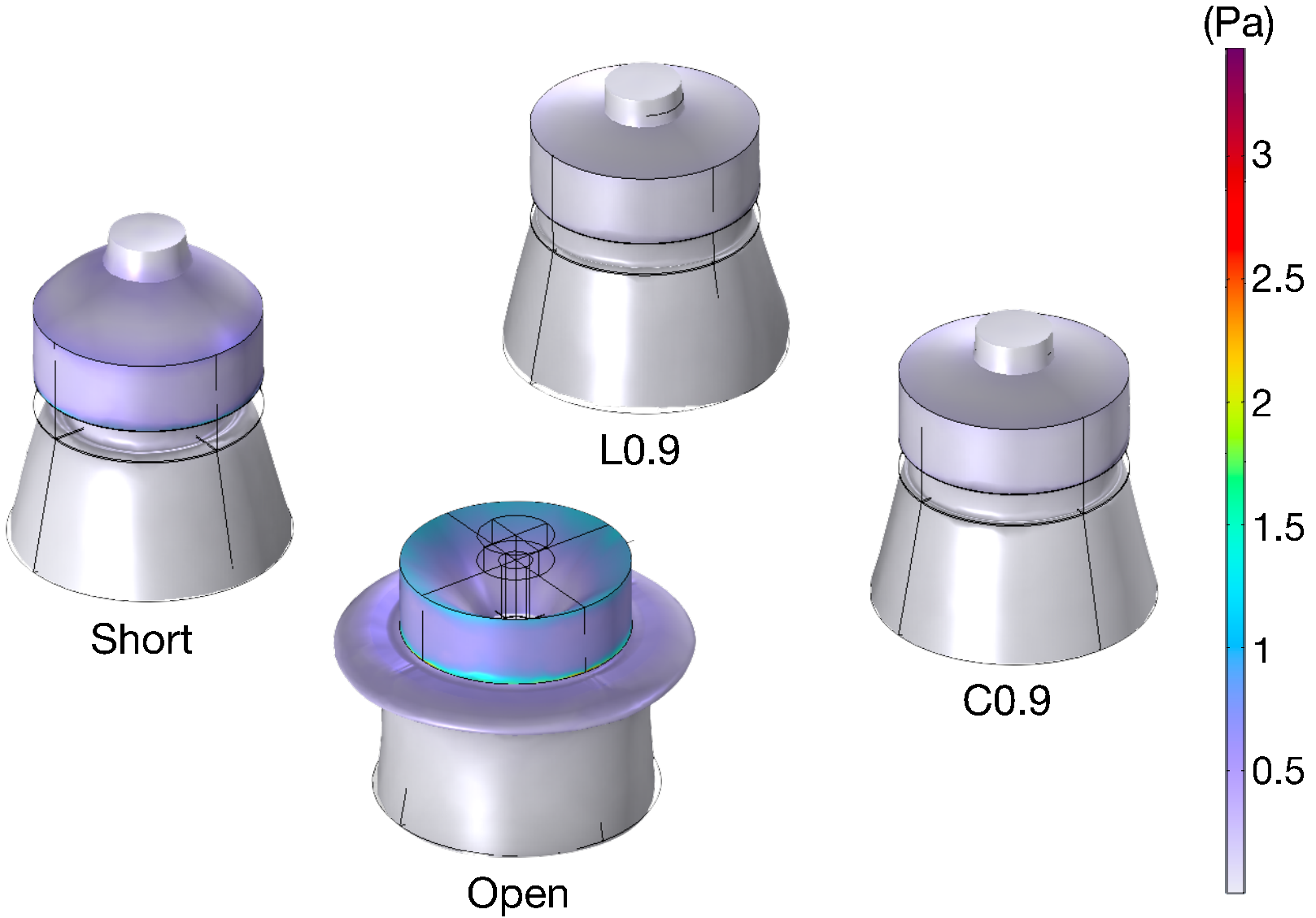}}
  \caption{Stress distributions of acoustic reflectors under various load conditions measured at 112\,$\mu$s.}\label{fig:Stress_112us}
\end{figure}

\textcolor{black}{Fig.~\!\ref{fig:Stress_112us} illustrates the stress distribution of acoustic reflectors measured under four distinct load conditions. It is evident that the stress profiles vary significantly across these conditions. As shown in the figure, at 112\,$\mu$s, the stress in the reflector with an open-circuit load reaches its negative peak, whereas the stress in the reflector with a short-circuit load reaches its positive peak. This observation is consistent with the theoretical analysis presented in Section~\!\ref{sec:MotiIQ}, which indicates that reflector coefficients generated by open and short circuits have identical amplitudes but opposite phases. In contrast, the stress in reflectors loaded with L0.9 and C0.9 is nearly zero at this time instant. This is because the reflector coefficients associated with inductive and capacitive loads exhibit phase shifts of 90$^{\circ}$ and $-$90$^{\circ}$, respectively, relative to the open circuit. Consequently, when the stress for the open-circuit load reaches its peak at 112\,$\mu$s, the stress for the inductive and capacitive load conditions crosses zero.}

\begin{figure}[htb]
\centerline{\includegraphics[width=8.5cm]{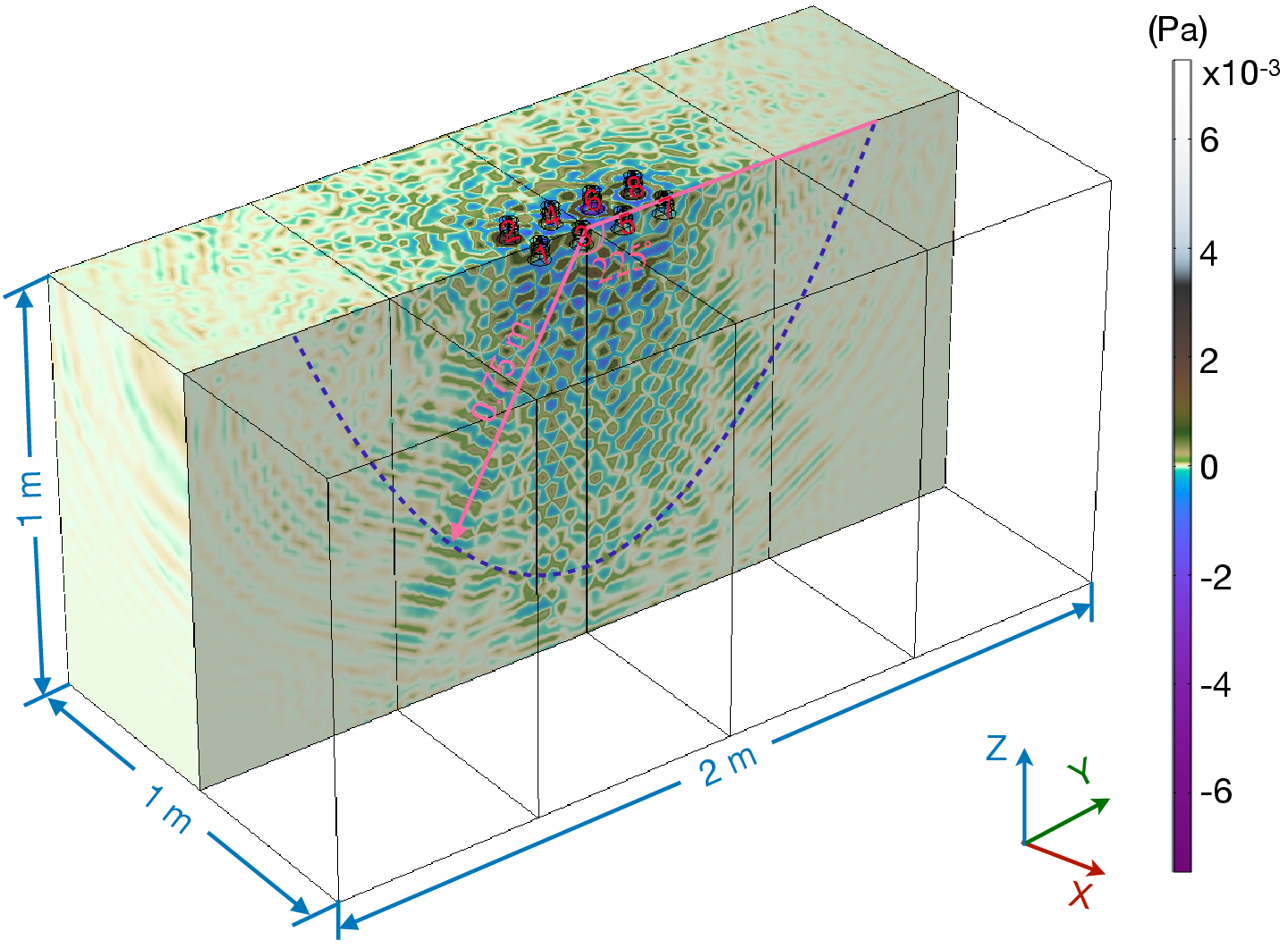}}
  \caption{Reflected beams generated by a UA-RIS with 8 acoustic reflectors using the synthetic reflection scheme.}\label{fig:3DBeam}
\end{figure} 

\textcolor{black}{Fig.~\!\ref{fig:3DBeam} illustrates the reflected beams generated by a UA-RIS comprising eight reflectors under a monochromatic plane wave excitation. The simulation environment is a water tank with dimensions of 1\,m$\times$1\,m$\times$2\,m, and the source signal operates at a frequency of 28\,kHz. Each reflector is labeled with a red identification number, and the spacing between adjacent reflectors is set to $2\lambda$, where $\lambda$ denotes the acoustic wavelength. To evaluate the acoustic directivity, 72 probes are uniformly placed along the blue dotted line on the Y-Z plane, each positioned 75\,cm from the center of the reflector array. Pairs of reflectors aligned along the X-axis are grouped for synthetic reflection, and their respective load impedances are provided in Table~\!\ref{tab:LdCon}. With this configuration, the UA-RIS is designed to produce a directional beam with maximum gain at 225$^{\circ}$ in the Y-Z plane.}

\begin{table}[htp]
\footnotesize
\centering
\setlength{\tabcolsep}{4pt} % reduced horizontal padding
\caption{Load impedance for different methods.}
\label{tab:LdCon}
\begin{tabular}{|c|c|c|c|c|c|c|c|c|}
\hline
\rowcolor[HTML]{EFEFEF} 
ID                                                   & 1                                                   & 2                                                   & 3                                                   & 4                            & 5                                                  & 6                             & 7                                                  & 8                            \\ \hline
\cellcolor[HTML]{EFEFEF}{\color[HTML]{000000} Syn.}  & \cellcolor[HTML]{E3DFDF}150\,$\Omega$ & \cellcolor[HTML]{C0C0C0}150\,$\Omega$ & \cellcolor[HTML]{E3DFDF}667\,$\Omega$ & \cellcolor[HTML]{C0C0C0}C06  & \cellcolor[HTML]{E3DFDF}18\,$\Omega$ & \cellcolor[HTML]{C0C0C0}L10   & \cellcolor[HTML]{E3DFDF}50\,$\Omega$ & \cellcolor[HTML]{C0C0C0}C10  \\ \hline
\cellcolor[HTML]{EFEFEF}{\color[HTML]{000000} 1-bit} & \cellcolor[HTML]{C0C0C0}Open                        & \cellcolor[HTML]{E3DFDF}Open                        & \cellcolor[HTML]{C0C0C0}Open                        & \cellcolor[HTML]{E3DFDF}Open & \cellcolor[HTML]{C0C0C0}Short                      & \cellcolor[HTML]{E3DFDF}Short & \cellcolor[HTML]{C0C0C0}Open                       & \cellcolor[HTML]{E3DFDF}Open \\ \hline
\end{tabular}
\end{table}

\textcolor{black}{As shown in Fig.~\!\ref{fig:3DBeam}, the reflected beam exhibits strong directivity at 225$^{\circ}$ as intended. Additionally, a grating lobe is observed near 240$^{\circ}$, resulting from the reflector spacing exceeding $1\lambda$. In the simulation, the absence of matching materials between the head mass and the surrounding water leads to a significant acoustic impedance mismatch at the interface, thereby reducing the amplitude of the reflected pressure signal. In practical implementations, this limitation can be addressed by applying multiple matching layers, an approach commonly employed in commercial acoustic transducers, to effectively minimize reflection losses at exposed boundaries.}

\begin{figure}[htb]
\centerline{\includegraphics[width=6.5cm]{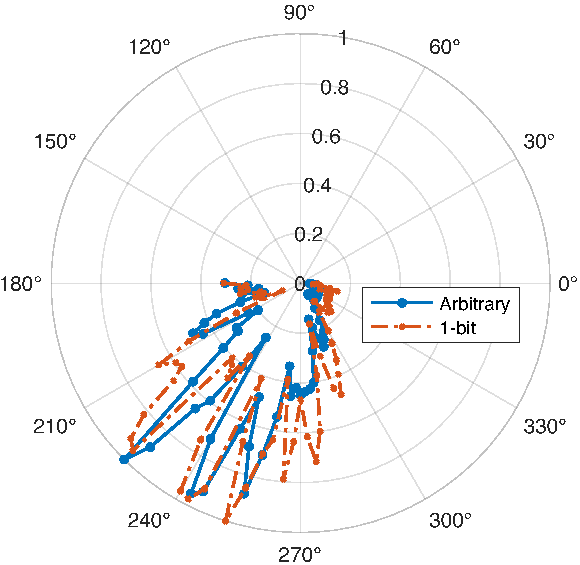}}
  \caption{Normalized reflected beam patterns produced by different schemes.}\label{fig:polar_arb_1bit}
\end{figure}

\textcolor{black}{In Fig.~\!\ref{fig:polar_arb_1bit}, we compare the reflected beam patterns generated by the proposed synthetic scheme and the conventional 1-bit coding scheme that is commonly used in RF-RIS. The beam amplitudes are normalized to their respective peak values. The simulation setup is identical to that in Fig.~\!\ref{fig:3DBeam}, and the load impedances applied to each reflector are listed in Table~\!\ref{tab:LdCon}.}

\textcolor{black}{As shown in Fig.~\!\ref{fig:polar_arb_1bit}, both schemes successfully steer the main beam toward 225$^{\circ}$ as expected. However, the synthetic scheme produces a noticeably narrower main lobe compared to the 1-bit coding approach, indicating improved beam directivity. Moreover, the side lobe levels produced by the synthetic scheme are significantly lower. For example, at 265$^{\circ}$, 275$^{\circ}$, and 290$^{\circ}$, the normalized amplitudes of side lobes under the synthetic scheme are 0.46, 0.43, and 0.27, respectively, whereas the corresponding values under the 1-bit scheme are 0.79, 0.72, and 0.48. These results demonstrate that the UA-RIS configured with the synthetic scheme achieves superior beamforming performance with enhanced directivity and reduced side lobe levels.}

%=================================================
\subsection {Tank Experiments}
\label{sec:TankExp}

\subsubsection{Experimental settings}
\begin{figure}[htb]
\centerline{\includegraphics[width=7.5cm]{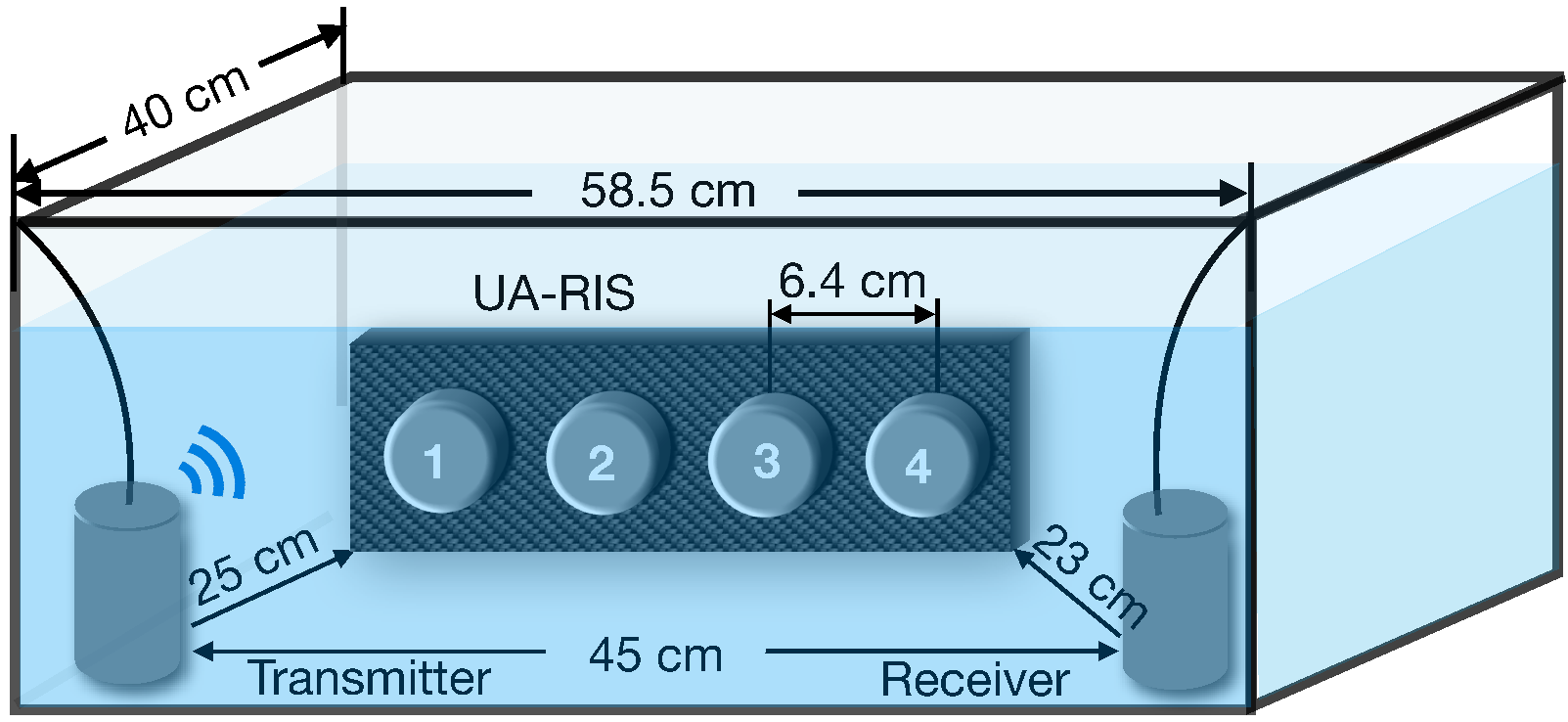}}
  \caption{Settings of tank tests.}\label{fig:TankScen}
\end{figure}

Before evaluating the performance of the entire system in the lake, we first verified the functionality of a single reflection unit in a tank environment. The experimental setup is shown in Fig.~\!\ref{fig:TankScen}, where two BTech Acoustic BT-2RCL~\cite{btech2024btech} omnidirectional acoustic transducers serve as the transmitter and receiver, respectively. The transmitter is driven by a Siglent 1062X arbitrary waveform generator, emitting a 28.23\,kHz continuous wave (CW) with a peak-to-peak voltage of 20.0\,V. The receiver is connected to a Siglent 1104X-E digital oscilloscope for data acquisition.

Four reflectors are placed at the bottom of the tank, with a spacing of 6.4\,cm (1.2 wavelengths at 28.23\,kHz) between neighboring reflectors. In the experiments, impedance matching between the transducer and the signal generator or oscilloscope was not performed, as our focus was on the relative changes in the received signal strength influenced by the reflector, rather than the absolute values.

\subsubsection{Resistive load modulation}
\begin{figure}[htb]
\centerline{\includegraphics[width=7.5cm]{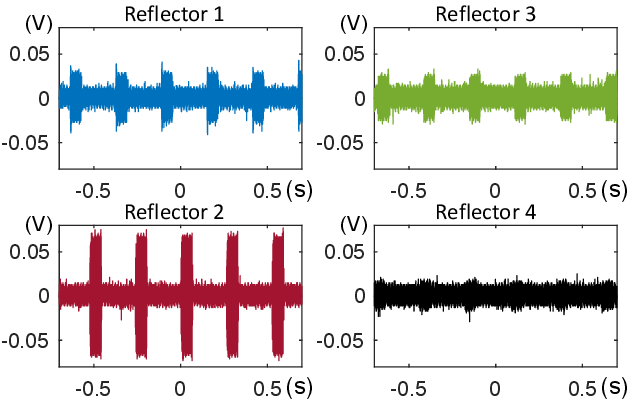}}
  \caption{Received signal strength with one selected reflector programmed for 1-bit coding at a time.}\label{fig:TankSingle}
\end{figure}

In Fig.~\!\ref{fig:TankSingle}, the MCU is programmed to perform 1-bit coding, periodically switching the load resistance of a specific reflector between short-circuit and open-circuit with durations of 60\,ms and 200\,ms, respectively. This configuration generates reflected waves with opposite phases. As depicted in the figure, altering the load resistance of the acoustic reflector remarkably affects the strength of the received signal. Using reflector 2 as an example, the voltage of the received signal is only 9\,mV with the short-circuit (0 load impedance). In contrast, when the circuit is open ($\infty$ load impedance), the received signal increases to 65\,mV, 7.2 times higher than in the short-circuit condition.

Additionally, Fig.~\!\ref{fig:TankSingle} demonstrates that changes in the reflector significantly affect the strength of the received signal. This is due to the proximity of the transmitter, receiver, and reflector, creating a near-field environment for acoustic propagation. Furthermore, the small size of the tank and its smooth surface contribute to rich multipath channels of the acoustic waves. With this setup, even slight adjustments to the reflector's position can have a dramatic impact on the received signal strength. For instance, when the load impedance of reflector 1 toggles between 0 and $\infty$, the received signal voltage shifts from 29\,mV to 13\,mV. In contrast, reflector 4 shows a much smaller change, with the voltage varying only between 19\,mV and 13\,mV, illustrating a less pronounced difference.

\begin{figure}[htb]
\centerline{\includegraphics[width=5.5cm]{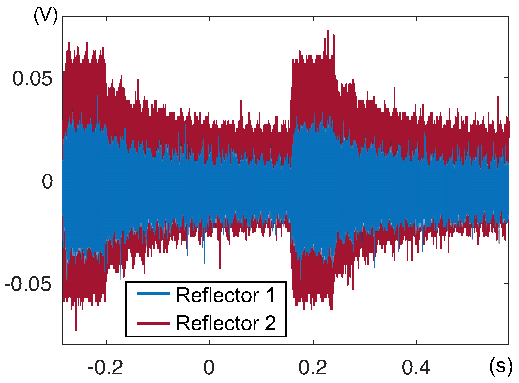}}
  \caption{Variation in strength of the received signal with changes in potentiometer's resistance.}\label{fig:TankPoten}
\end{figure}

In Fig.~\!\ref{fig:TankPoten}, the resistive network is connected to the reflector, and then the MCU adjusts the potentiometer's resistance to vary from 50\,$\Omega$ to 1.96\,k$\Omega$ in 200\,$\Omega$ increments. The results in the figure demonstrate that the digital potentiometer can effectively controls the amplitude of the reflected wave, enabling precise adjustments to the received signal strength. Using reflector 2 as an example, when the load resistance is set to 50\,$\Omega$, the reflected waves simulate a short-circuit condition, resulting in a relatively high signal strength of 63\,mV. As the resistance increases, the reflected wave gradually shifts toward an open-circuit situation, reducing the signal strength to 23\,mV. At the maximum resistance of 50\,k$\Omega$, supported by the AD8403AR50 potentiometer, the strength of the received signal can further decrease to approximately 9\,mV.

In the tank experiments, the received signals consist of a mixture of three components: (a) the direct wave, (b) waves reflected by the reflector one or more times, and (c) other multipath signals that do not interact with the reflector. The results presented in Fig.~\!\ref{fig:TankSingle} and Fig.~\!\ref{fig:TankPoten} reflect the combined effects of these three components. An intriguing question arises: how does a UA-RIS contribute to the received signal in an unbounded space? In other words, we aim to isolate component (b) to analyze its specific contribution to enhancing the strength of the received signal.

To achieve the above objective, let $s(t)$ and $r(t)$ represent the source and received signals at time $t$, respectively. Taking multipath effects into account, when the reflector's load states are in open-circuit and short-circuit, the corresponding received signals, which are denoted by $r_{op}(t)$ and $r_{sh}(t)$, respectively, can be described as follows:
\begin{equation}\label{eq:faou}
    \left\{
    \begin{array}{lll}
        \vspace{0.2cm}
        \!\!r_{op}(t)\!=\! \displaystyle\sum^N_{i=1}A_i e^{j\psi_i}s(t-\tau_i)\!+\!\displaystyle\sum^M_{k=1}B_ke^{j\phi_k} s(t-\tau_k),\\
        \vspace{0.13cm}
        \!\!r_{sh}(t)\!=\! \displaystyle\sum^N_{i=1}A_i e^{j\psi_i} s(t-\tau_i)\!+\!\displaystyle\sum^M_{k=1}B^\prime_k e^{j\phi^\prime_k} s(t-\tau_k),\\
    \end{array}
    \right.
\end{equation}
where $A_i$ and $\psi_i$ are the amplitude and phase of the channel coefficient for the direct wave and the multipath signals that do not interact with the reflector. $B_k$ and $\phi_k$ are the amplitude and phase of the channel coefficient for waves reflected by the reflector with infinite load impedance (open-circuit); $B^\prime_k$ and $\phi^\prime_k$ are that with zero load impedance (short-circuit).

In the second term of (\ref{eq:faou}), we restrict our consideration to waves reflected only once by the reflector since waves reflected multiple times are greatly attenuated and thus contribute minimally to the outcomes. With this restriction, $B^\prime_ke^{j\phi^\prime_k}\!=\!-B_ke^{j\phi_k}$ as the amplitudes of signals reflected by the reflector with 0 and $\infty$ load impedances are equivalent, though their phases are opposite. Consequently, for $[r_{op}(t) - r_{sh}(t)]/2$, only the waves once reflected by the reflector remain. Conversely, for $[r_{op}(t) + r_{sh}(t)]/2$, only the first term of the equations persists, representing the signal unaffected by the acoustic reflector. Subsequently, we will demonstrate the waveform of these terms through experimental results.

\begin{figure}[htb]
\centerline{\includegraphics[width=8.0cm]{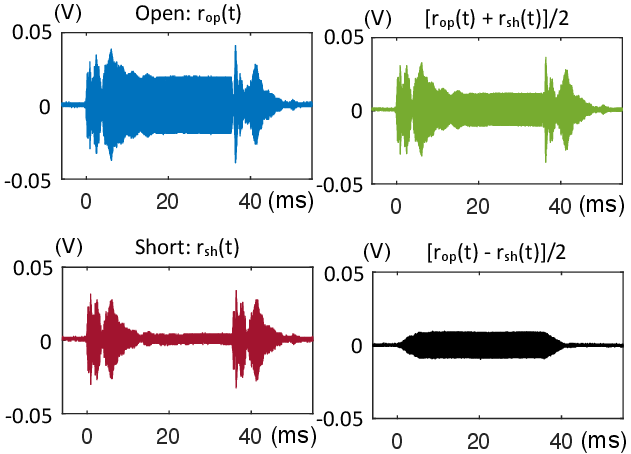}}
  \caption{Waveform comparison of received signals with and without the reflector.}\label{fig:TankPro}
\end{figure}

In Fig.~\!\ref{fig:TankPro}, the waveform generator emits a burst sinusoidal signal consisting of 1500 cycles. The peak-to-peak voltage and frequency of the source signal are set at 20\,V and 28.23\,kHz, respectively, identical to those presented in Fig.~\!\ref{fig:TankSingle}. As illustrated in the figure, due to the multipath effect, the beginnings and ends of $r_{op}$ and $r_{sh}$ exhibit significant fluctuations lasting approximately 15\,ms. Once all multipath signals have arrived at the receiver, the combined signal becomes stable. 

From Fig.~\!\ref{fig:TankPro}, it can be observed that the wave generated by the acoustic reflector, i.e., $[r_{op}(t) - r_{sh}(t)]/2$, is smooth. This indicates that the channel environment is consistent and that signals $r_{op}$ and $r_{sh}$ collected at different times are well aligned, ensuring the validity of subsequent analysis. By comparing the waveforms of $[r_{op}(t) + r_{sh}(t)]/2$ and $[r_{op}(t) - r_{sh}(t)]/2$, we have that the amplitude of the received signal created by a single reflector reaches 8.6\,mV, which is comparable to that generated by the sum of signals unaffected by the acoustic reflector. This result demonstrates the effectiveness of the constructed acoustic reflector and load network in manipulating the strength of the received signal.

\subsubsection{Reactive load modulation}
\begin{figure}[htb]
\centerline{\includegraphics[width=8.7cm]{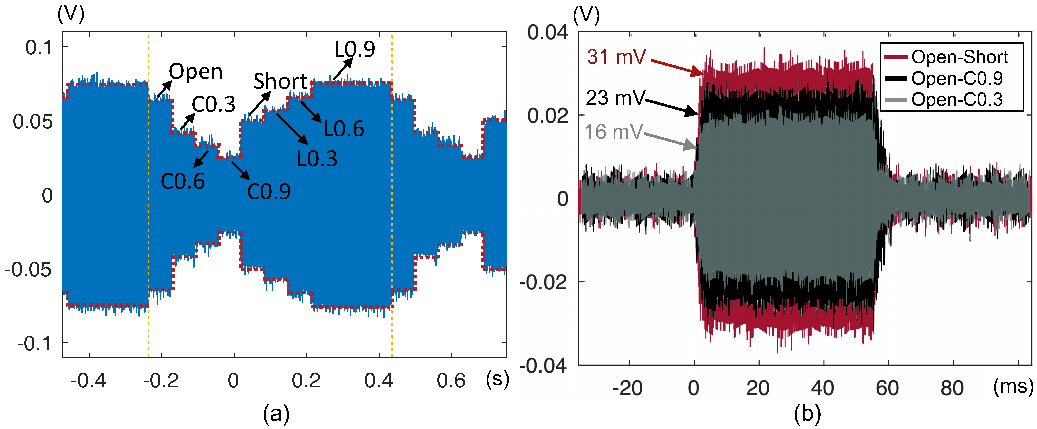}}
  \caption{Waveform of the received signal with variations in load impedance. (a) Original waveform. (b) Differential waveforms.}\label{fig:MultiCha}
\end{figure}

The multipath model presented in (\ref{eq:faou}) can also be used to validate the reflector's ability to produce the quadrature component for \textcolor{black}{synthetic reflection}. This can be achieved by programming the MCU to sequentially connect the subnetwork shown in Fig.~\!\ref{fig:Circuit} with the acoustic reflector. The waveform of the received signal is depicted in Fig.~\!\ref{fig:MultiCha}\,(a)\footnote{The experiment in Fig.~\!\ref{fig:MultiCha} was conducted on a different day than those in Figs.~\!\ref{fig:TankSingle} to \ref{fig:TankPro}. This difference resulted in varying acoustic environments and channel states, indicating that this experiment should be considered independent.}, where the source is 28.23\,kHz CW with 20\,V peak-to-peak voltage. In the figure, labels such as C0.3 and L0.9 indicate that the acoustic reflector is connected to the capacitive and inductive load network with reflection coefficients of 0.3 and 0.9 amplitude, respectively.

As depicted in Fig.~\!\ref{fig:MultiCha}\,(a), the received signal strength varies markedly when different reactive loads are connected to the reflector. For example, with a capacitive load labeled C0.9, the amplitude of the received signal is only 23\,mV. However, when switching to an inductive load labeled L0.9, the amplitude of the received wave increases to 72\,mV, which is 3.1 times greater than that observed with the previous capacitive load.

The results in Fig.~\!\ref{fig:MultiCha}\,(a) illustrate only the amplitude of the received signal influenced by the multipath effect. To accurately evaluate the phase and magnitude of the reflected wave generated by each reactive load, we need to further apply the approach outlined in Fig.~\!\ref{fig:TankPro}. In this approach, the waveform generator transmits a burst sinusoidal signal consisting of 1500 cycles at 28.23\,kHz with a 20\,V peak-to-peak voltage, followed by calculating the strength of the differential signals produced by the reflector with various load impedances.

Let $r_{c.9}$, and $r_{c.3}$ represent the received signals corresponding to the load conditions of the acoustic reflector being at C0.9, and C0.3, respectively. As outlined in Section~\!\ref{sec:StruAco}, the reflection coefficients for the reflector with load impedances $r_{op}$, $r_{sh}$, $r_{c.9}$, and $r_{c.3}$ are $1$, $-1$, $0.9e^{-j\frac{2}{\pi}}$, and $0.3e^{-j\frac{2}{\pi}}$, respectively. Consequently, based on (\ref{eq:faou}), we have that
\begin{equation}\label{eq:oqqw}
    \left\{
    \begin{array}{lll}
        \vspace{0.2cm}
        r_{op}(t)\!-r_{sh}(t)&\!\!\!\!\!=\!\!\!\!\!\!&2\displaystyle\sum^M_{k=1}\!B_k\,e^{j\phi_k} s(t\!-\!\tau_k),\\
        \vspace{0.13cm}
        r_{op}(t)\!-r_{c.9}(t)&\!\!\!\!\!=\!\!\!\!\!\!&\left(1\!-\!0.9\,e^{-\frac{j\pi}{2}}\right)\!\displaystyle\sum^M_{k=1}\!B_k\,e^{j\phi_k}s(t\!-\!\tau_k),\\
        \vspace{0.13cm}
        r_{op}(t)\!-r_{c.3}(t)&\!\!\!\!\!=\!\!\!\!\!\!&\left(1\!-\!0.3\,e^{-\frac{j\pi}{2}}\right)\!\displaystyle\sum^M_{k=1}\!B_k\,e^{j\phi_k}s(t\!-\!\tau_k).\\
    \end{array}
    \right.
\end{equation}
According to (\ref{eq:oqqw}), the following ratio can be obtained:
\begin{equation}\label{eq:a14v}
    \begin{aligned}
       & \abs{r_{op}(t)\!-r_{sh}(t)}:\abs{r_{op}(t)\!-r_{c.9}(t)}:\abs{r_{op}(t)\!-r_{c.3}(t)} \\
       =&\; 2: \abs{1\!-\!0.9\,e^{-\frac{j\pi}{2}}}: \abs{1\!-\!0.3\,e^{-\frac{j\pi}{2}}}\\
       =&\; 1: 0.68: 0.52.
    \end{aligned}
\end{equation}

Now, we can use the above ratio to check the accuracy of the quadrature component generated by the reflector. As shown in Fig.~\!\ref{fig:MultiCha}\,(b), the average strengths of differential signals $r_{op}-r_{sh}$, $r_{op}-r_{c.9}$, and $r_{op}-r_{c.3}$ are 31\,mV, 23\,mV, and 16\,mV, respectively; the ratio of these values is 1: 0.74: 0.52, close to the idea case calculated in (\ref{eq:a14v}). This validates the ability of the constructed reflector to generate the negative quadrature component for \textcolor{black}{synthetic reflection}.

\begin{figure}[htb]
\centerline{\includegraphics[width=6.5cm]{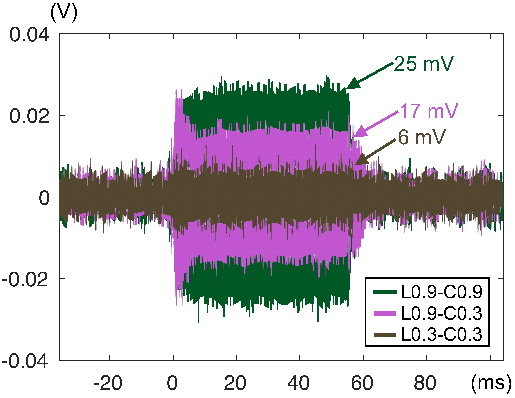}}
  \caption{Differential signal with inductive and capacitive loads at the acoustic reflector.}\label{fig:MultiCha1}
\end{figure}

We also checked the differential waves with inductive load at the reflector using the following ratio:
\begin{equation}\label{eq:mnut}
    \begin{aligned}
       & \abs{r_{l.9}(t)\!-r_{c.9}(t)}\!:\!\abs{r_{l.9}(t)\!-r_{c.3}(t)}\!:\!\abs{r_{l.3}(t)\!-r_{c.3}(t)} \\
       =&0.9\!\times\!\!\abs{e^{\frac{j\pi}{2}}\!\!-\!e^{-\frac{j\pi}{2}}}\!:\! 0.3\!\times\!\!\abs{0.3\,e^{\frac{j\pi}{2}}\!\!-\!\,e^{-\frac{j\pi}{2}}}\!:\! 0.3\!\times\!\!\abs{e^{\frac{j\pi}{2}}\!\!-\!e^{-\frac{j\pi}{2}}}\\
       =&\; 1: 0.67: 0.33,
    \end{aligned}
\end{equation}
where $r_{l.9}$ and $r_{l.3}$ represent the received signals corresponding to the load conditions of the acoustic reflector being in L0.9 and L0.3, respectively.

As illustrated in Fig.~\!\ref{fig:MultiCha1}, the average strengths of differential signals $r_{l.9}-r_{c.9}$, $r_{l.9}-r_{c.3}$, and $r_{l.3}-r_{c.3}$ are 25\,mV, 17\,mV, and 6\,mV, respectively; the ratio of these values is 1: 0.68: 0.24, close to the idea case calculated in (\ref{eq:mnut}). This result validates the ability of the constructed reflector to generate the positive quadrature component for \textcolor{black}{synthetic reflection}.

\textcolor{black}{In the tank tests, the relatively strong reflections caused by the tightly spaced setup may lead to an overestimation of reflector gains in more complex or challenging environments. To address this limitation, lake tests are conducted in Section~\!\ref{sec:LakTes} to further evaluate the performance of UA-RIS in a real-world scenario with dynamic channel conditions.}

%=================================================
\subsection {Lake Tests}
\label{sec:LakTes}

\subsubsection{Experimental setup}
\begin{figure}[htb]
\centerline{\includegraphics[width=9.0cm]{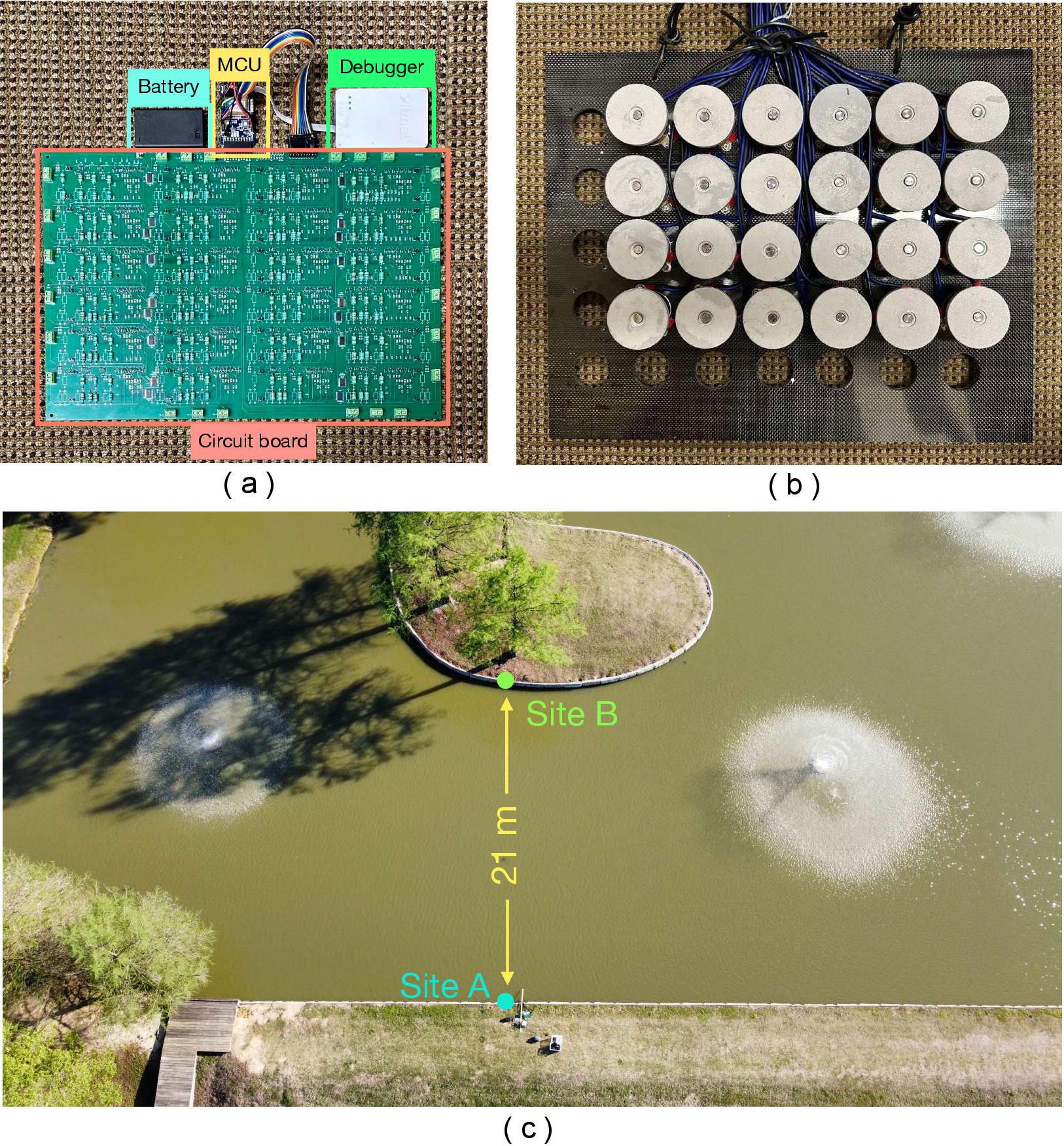}}
  \caption{Lake experiments. (a) 24-channel control board. (b) Acoustic reflection units. (c) Experimental setup.}\label{fig:BoardPro}
\end{figure}

We conducted lake experiments to validate the functionality of our UA-RIS system. The control board and the reflecting array are depicted in Fig.~\!\ref{fig:BoardPro}\,(a) and Fig.~\!\ref{fig:BoardPro}\,(b), respectively, while the experimental setup is illustrated in Fig.~\!\ref{fig:BoardPro}\,(c). In the test, the arbitrary waveform generator emitted a 28\,kHz CW signal at 15\,dBm. This signal was then amplified using a Fosi M03 audio power amplifier to drive the BTech omnidirectional acoustic transducer. At the receiving end, another BTech transducer positioned 21\,m away from the transmitter captured the acoustic signal. The received waveform was initially amplified by a Sound Devices MM-1 microphone amplifier, which provided a 54\,dB gain, and then recorded with a PicoScope 4224 portable oscilloscope.

\subsubsection{UA-RIS close to sender}
As shown in Fig.~\!\ref{fig:BoardPro}\,(c), in the initial lake test, the receiver was deployed at site B , while the transmitter and the UA-RIS were located at site A, with the latter positioned 15 wavelengths behind the former. From the receiver's perspective, this setup emulates a scenario where an acoustic signal from a virtual source travels a certain distance before being reflected by a UA-RIS positioned 21 meters away. By adjusting the emission power of the signal generator or the gain of the power amplifier, we can replicate different distances of signal propagation between the virtual source and the UA-RIS.

\begin{figure}[htb]
\centerline{\includegraphics[width=7.5cm]{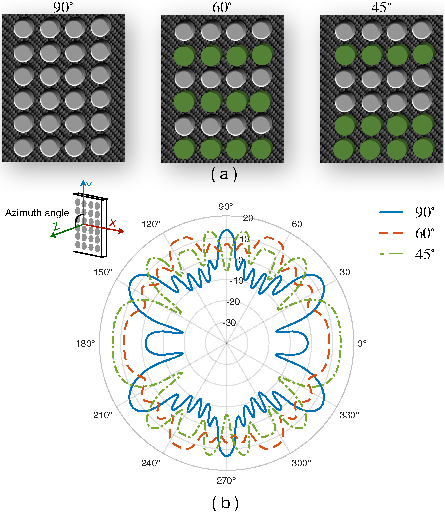}}
  \caption{Load configuration of UA-RIS for beamforming at the transmitter side. (a) Coding schemes for steering. (b) Azimuth beam pattern of the reflected waves.}
  \label{fig:1bitCoding}
\end{figure}

The diagrams of the three load configurations, i.e., the coding schemes used in the first test, are depicted in Fig.~\!\ref{fig:1bitCoding}\,(a). In these diagrams, the initial loads of the gray and green reflectors are set to open and short-circuits, respectively. The coordinate system and the azimuth beam patterns generated by these three coding schemes are illustrated in Fig.~\!\ref{fig:1bitCoding}\,(b). \textcolor{black}{In the figure, the beam patterns are numerically derived by processing the reflection patterns corresponding to different RIS configurations. Each configuration corresponds to a fixed incident angle ($90^\circ$, $60^\circ$, and $45^\circ$), and the reflected field is synthesized by correlating the phase-coded reflection sequences with theoretical steering vectors over a range of azimuth angles in the X-Y plane. The resulting beam patterns shown in Fig.~\!\ref{fig:1bitCoding}\,(b) are computed based on this correlation to indicate the directional gain for each configuration.}

\begin{figure}[htb]
\centerline{\includegraphics[width=6.5cm]{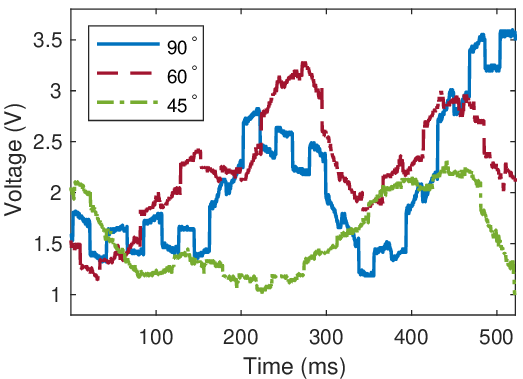}}
  \caption{Envelopes of received signals with various coding schemes implemented at UA-RIS.}\label{fig:LakeTest1}
\end{figure}

In Fig.~\ref{fig:LakeTest1}, the UA-RIS toggles the load of each reflector between short-circuit and open-circuit states every 20\,ms, following the initial load coding scheme illustrated in Fig.~\ref{fig:1bitCoding}\,(a). This configuration results in varying levels of received signal strength, influenced by the beam pattern generated by UA-RIS. Notably, when the azimuth angle of the main beam is set to 90$^{\circ}$, the reflected waves align directly with the receiver's direction. This alignment leads to amplitude variations averaging 280\,mV and peaking at up to 400\,mV as the reflector's load cycles between the two states. In contrast, when the beam's azimuth angles are adjusted to 60$^{\circ}$ and 45$^{\circ}$, the differential voltage at the receiver decreases markedly to 100\,mV and 33\,mV, respectively. This reduction is due to the main lobe of the reflected waves deviating from the receiver's direct path, which diminishes wave superposition at the receiver. Consequently, this lessens the impact of the reflected wave on the received signal strength.

Based on Fig.~\!\ref{fig:LakeTest1}, we can deduce the potential enhancement in data rate achievable with the UA-RIS. As depicted in the figure, directing the main beam to 90$^{\circ}$ elevates the received signal from 1\,V (blue line at 320\,ms) to an average of 1.28\,V, yielding a 2.15\,dB increase in SNR. Under these conditions, employing frequency-shift keying (FSK) or phase-shift keying (PSK) modulations could increase the data rate by approximately 63.8\% while maintaining the same BER. In the optimal scenario, with a 400\,mV increase in the received signal, the SNR improvement reaches 2.9\,dB, potentially boosting the data rate by up to 96\% compared to configurations without UA-RIS.

Now, let's assess the impact of UA-RIS on communication range. Suppose the initial communication distance to be 0.5\,km. If the received voltage increases from 1\,V to 1.4\,V, this corresponds to a gain of 3\,dB in signal pressure at the receiver side. Given these parameters, we can derive the following equation:
\begin{equation}
\label{eq:8aqh}
  10\,\alpha\left(\log_{10}R_y-\log_{10}R_x\right) +\beta \left(R_y-R_x\right) = \Delta \text{SNR},
  \end{equation}
where $R_x\!=\!0.5$\,km represents the original communication distance. $R_y$ denotes the extended communication range achieved with the support of UA-RIS. $\Delta \text{SNR}\!=\!2.9$\,dB is the received SNR gain achieved by the UA-RIS. The spreading coefficient, denoted by $\alpha$, is 1 for cylindrical waves and 2 for spherical waves. The coefficient $\beta$ represents the absorption loss of acoustic waves. According to the Francois-Garrison model\cite{lurton2002introduction}, $\beta\!=\!6.1$\,dB/km for a 28\,kHz acoustic signal in seawater under typical conditions (10\textdegree C temperature, 35 ppt salinity, and pH 8).

Referring to (\ref{eq:8aqh}), we have that in deep water environments, where $\alpha\!=\!2$ for spherical spreading, $R_y$ can reach 0.64\,km. With the support of UA-RIS, the communication range is extended by 28\%. In shallow water environments, characterized by $\alpha\!=\!1$ for cylindrical spreading, $R_y$ further extends to 0.73\,km, marking a 46\% enhancement from the original range.

\subsubsection{UA-RIS close to receiver}
\begin{figure}[htb]
\centerline{\includegraphics[width=7.5cm]{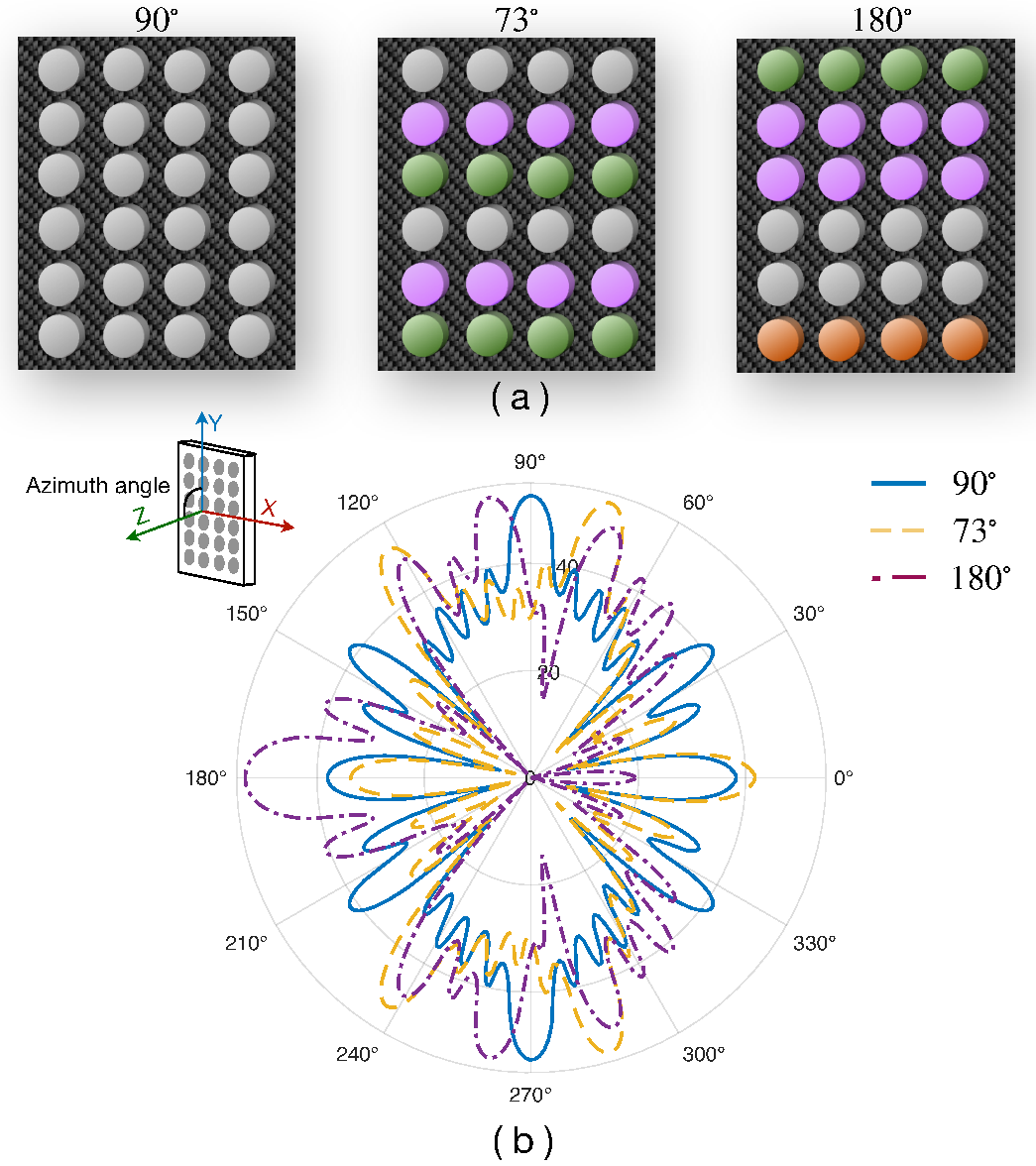}}
  \caption{Load configuration of UA-RIS for beamforming at the receiver side. (a) Coding schemes for steering. (b) Azimuth beam pattern of the reflected waves.}\label{fig:2bitCoding}
\end{figure}

In the second lake test, the transmitter and receiver remained in their original positions, while the UA-RIS was moved to site B, 15 wavelengths behind the receiver. To evaluate the UA-RIS's performance with \textcolor{black}{synthetic reflection}, three different load coding schemes were applied. As illustrated in Fig.~\!\ref{fig:2bitCoding}\,(a), the initial load settings for the gray, green, purple, and orange reflectors were configured to open, short, L0.9, and C0.9, respectively. The azimuth beam patterns resulting from these coding schemes are shown in Fig.~\!\ref{fig:2bitCoding}\,(b), where the main lobes of the reflected beams are directed at 90$^{\circ}$, 73$^{\circ}$, and 180$^{\circ}$, respectively.

\begin{figure}[htb]
\centerline{\includegraphics[width=6.5cm]{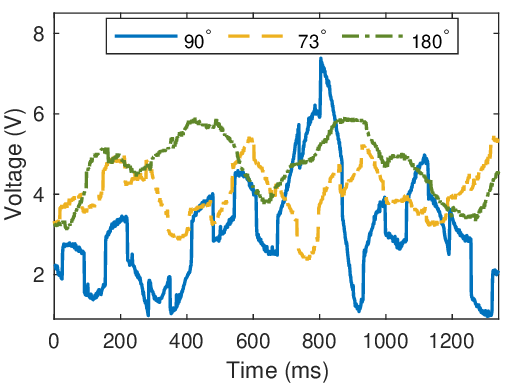}}
  \caption{Envelopes of received signals with various coding schemes implemented at UA-RIS.}\label{fig:LakeTest1019}
\end{figure}

In Fig.~\!\ref{fig:LakeTest1019}, the UA-RIS alternates the load of each reflector between short-circuit and open-circuit or L0.9 and C0.9 every 66\,ms, following the initial coding scheme depicted in Fig.~\!\ref{fig:2bitCoding}\,(a). From the figure, significant variations in the intensity of the received signal can be observed, which are dependent on the beam pattern generated by the UA-RIS.

When the main lobe of the reflected beam is directed at 90$^{\circ}$ towards the receiver, the strength of the received signal is considerably enhanced. Specifically, as the reflector's load cycles between the two states, the average amplitude variation is 1.04\,V, with the maximum variation reaching up to 1.4\,V. In contrast, when the beam is steered to 73$^{\circ}$ and 180$^{\circ}$, the main lobe of the reflected wave deviates from the receiver's direction, resulting in reduced average amplitude variations of 0.51\,V and 0.08\,V, respectively.

Referring to Fig.~\!\ref{fig:LakeTest1019}, the implementation of the UA-RIS on the receiver side can significantly enhances both the data rate or the communication range. When the beam of the reflected wave targets the receiver, the UA-RIS elevates the voltage of the received signal from 1.75\,V to 2.83\,V (as indicated by the blue line at 150\,ms), yielding a maximum SNR gain of approximately 4.2\,dB. This enhancement allows for a potential increase in the data rate by 163\% for both FSK and BPSK modulation schemes, while maintaining the same BER. On average, the UA-RIS boosts the received signal voltage from 3.04\,V to 4.08\,V, resulting in an SNR gain of 2.56\,dB, and facilitating an average of 80.3\% data rate increment.

Considering an initial communication distance of 0.5\,km between the transceiver units, and applying (\ref{eq:8aqh}), the communication distance in a deep-water environment can be extended to 0.7\,km with the 4.2\,dB SNR gain, translating to an approximate 40.6\% increase in range while preserving the same BER. In shallow water scenarios, characterized by a spreading coefficient of 1, the communication distance can be further extended to 0.83\,km, thus augmenting the communication range by about 66\%.

\textcolor{black}{
\subsubsection{Discussions}
In the lake tests, using CW signals primarily validate steady-state characteristics such as beam steering and reflection capabilities of UA-RIS, which served as foundational assessments for this work. Practical modulation schemes that commonly used for underwater acoustic communications, such as frequency shift keying (FSK), phase shift keying, and orthogonal frequency division multiplexing (OFDM) will introduce additional complexities and transient behaviors not fully captured by CW testing. However, preliminary analytical and experimental results collected in the field test provide strong evidence that performance improvements are expected when transitioning to practical modulation schemes.}

\textcolor{black}{For precise beamforming, the UA-RIS needs to be aware of its own attitude, which may vary due to ocean currents. To ensure optimal performance, real-time attitude estimation can be achieved by incorporating data from an inertial measurement unit (IMU) and a pressure sensor. For instance, the gyroscope within the IMU provides angular velocity measurements that enable tracking of three-dimensional orientation changes over time, while the pressure sensor supplies depth information essential for vertical positioning.}

\textcolor{black}{Our proof-of-concept experiments are based on a fixed, known geometry and therefore do not incorporate incident-angle estimation or transceiver localization. Accurate estimation of the angle of incident waves for steering the reflected signal in the desired direction is essential yet challenging tasks. To solve this problem, UA-RIS can leverage the spatial characteristics of the received signal across the reflection elements to estimate the angle of arrival (AoA) of an incident signal. In scenarios where the RIS is equipped with sensing capabilities either through embedded low-power sensors or via hybrid active elements, it becomes possible to treat the RIS as an antenna array that samples the incoming wavefront. By modeling the received signal as a superposition of array response vectors parameterized by the incident angle, traditional array signal processing techniques such as MUSIC (multiple signal classification) or ESPRIT (estimation of signal parameters via rotational invariance techniques) can be applied to estimate the AoA with high resolution. This approach allows precise directional information to be extracted, which is critical for dynamically configuring the RIS reflection pattern to optimize signal propagation and enhance communication performance.}

%%=========================================
%%=========================================

\section{Conclusion and Future Work}
\label{sec:Con}
In this paper, we explored the design of UA-RIS to enhance the range and speed of underwater acoustic communications. We addressed several challenging issues critical to system implementation, including  wide dynamic range and low frequency of acoustic signals. To optimize performance, the proposed UA-RIS strategically aligns reflection units based on the angles of incident waves and their intended directions, utilizing \textcolor{black}{synthetic reflection} method to shape the reflected waves. This approach can precise control over the phase and amplitude of the signals, enabling advanced beam-steering capabilities.

Results from \textcolor{black}{COMSOL Multiphysics simulation and} tank experiments  demonstrated the capability of acoustic reflectors to finely tune the phase and amplitude of signals by adjusting the load impedance. Additionally, we evaluated a UA-RIS setup comprising 24 reflection units during lake trials, which proved effective in significantly enhancing or diminishing signal strength in designated directions. This adjustment improved receiving SNR and minimized interference at specific locations. Experimental results \textcolor{black}{using a CW signal as the source} indicate that deploying the UA-RIS near the transmitter can extend the communication range by up to 28\% and 46\% in deep water and shallow water environments, respectively. \textcolor{black}{When the distance between communicating parties is fixed, the UA-RIS can enhance the received SNR by an average of 2.13\,dB, with peak improvements reaching up to 2.92\,dB in certain cases.} In addition, moving the UA-RIS near the receiver extends the communication range in both deep and shallow water environments by 40.6\% and 66\%, respectively. Furthermore, with the distance fixed, situating the UA-RIS near the receiver not only improves \textcolor{black}{SNR by an average of 2.56\,dB but can also achieve increases of up to 4.2\,dB} under specific conditions.

UA-RIS technology is still in its early stages, its real-world deployment requires further development. In particular, integrating this new technology into existing systems necessitates critical modifications to current medium access control protocols. Adapting these protocols \textcolor{black}{with different modulation schemes (e.g., FSK, PSK, and OFDM)} will enable the seamless integration of UA-RIS into established underwater network infrastructures, a topic we intend to investigate in our future research.

%%=========================================
%%=========================================

\bibliographystyle{IEEEtran}
%\bibliography{UnderRIS}


\begin{IEEEbiography}[{\includegraphics[width=26mm,clip,keepaspectratio]{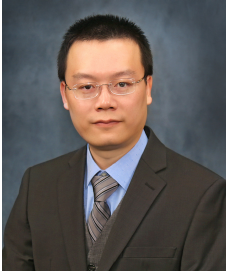}}]
{Dr. Yu Luo} received the B.S. degree and the M.S. degree in electrical engineering from the Northwestern Polytechnical University, China, in 2009 and 2012, respectively. In 2015, he received the Ph.D. degree in computer science and engineering from University of Connecticut, Storrs. Dr. Luo is currently an Associate Professor at Mississippi State University. His major research focus on the IoT, Edge Computing, RF energy harvesting hardware, security in RF energy harvesting wireless networks, and underwater wireless networks. He is a Co-recipient of the Best Paper Award in IFIP Networking 2013 and Chinacom 2016.
\end{IEEEbiography}

\begin{IEEEbiography}[\vspace{-0.2cm}{\includegraphics[width=26mm,clip,keepaspectratio]{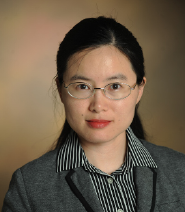}}]
{Dr. Lina Pu} received the B.S. degree in electrical engineering from the Northwestern Polytechnical University, Xi'an, China in 2009 and the Ph.D. degree in Computer Science and Engineering from University of Connecticut, Storrs. Dr. Pu is currently an Assistant Professor at University of Alabama. Her research interests lie in the area of edge computing, RF energy harvesting wireless networks, security in the sustainable IoT, and underwater acoustic networks. She owned IFIP Networking 2013 best paper award.
\end{IEEEbiography}

\begin{IEEEbiography}[\vspace{0.0cm}{\includegraphics[width=26mm,clip,keepaspectratio]{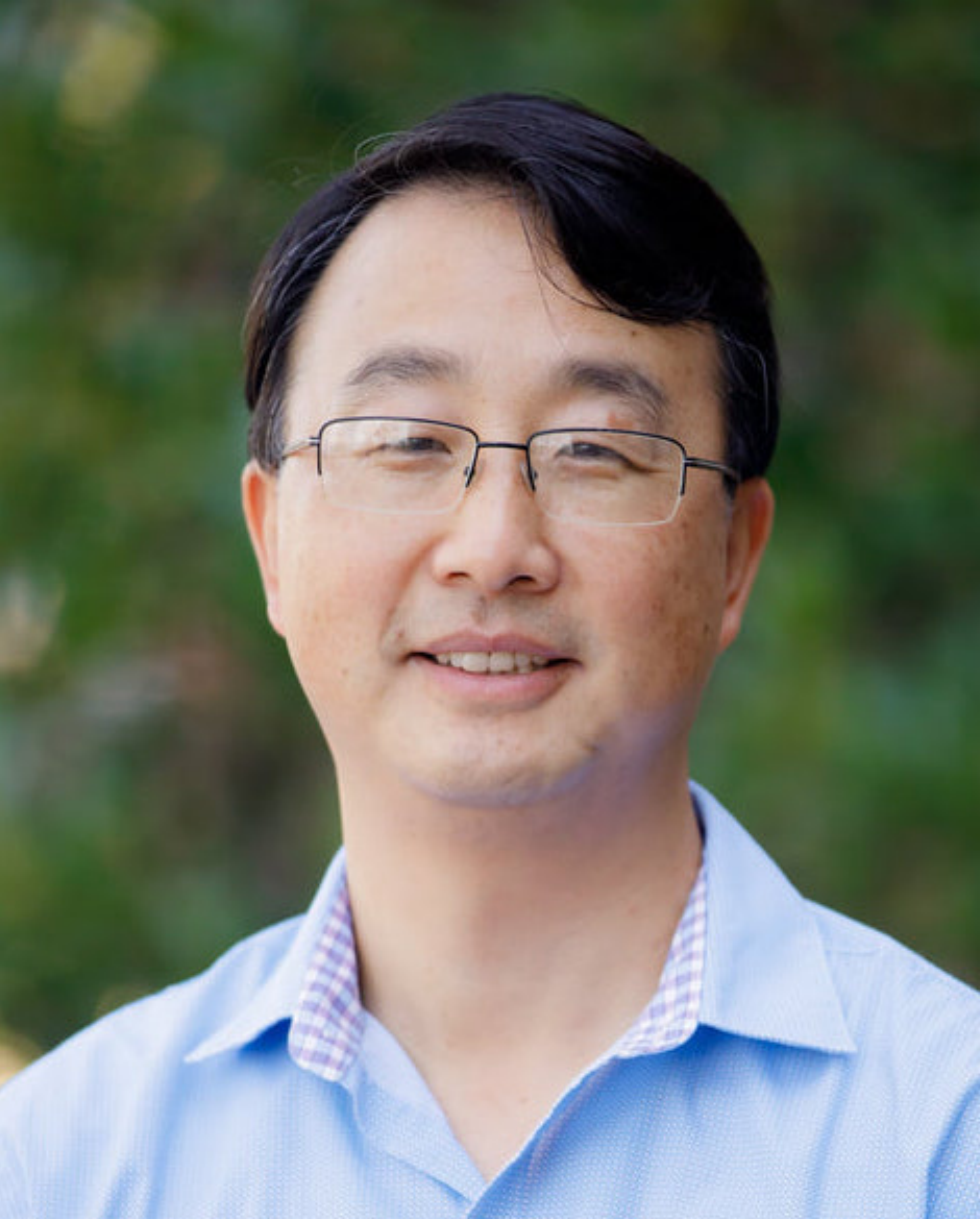}}]
{Dr. Aijun Song} (Senior Member, IEEE) received the Ph.D. degree in electrical engineering from the University of Delaware, Newark, DE, USA, in 2005. From 2005 to 2008, he was a Postdoctoral Research Associate with the College of Earth, Ocean, and Environment, University of Delaware. During this period, he was also an Office of Naval Research (ONR) Postdoctoral Fellow, supported by the Special Research Award in the ONR Ocean Acoustics program. From 2008 to 2015, he was a Research Professor with the University of Delaware. He is
currently an Associate Professor with the Department of Electrical and Computer Engineering, University of Alabama, Tuscaloosa, AL, USA. His research interests include underwater wireless communications and networking, underwater robotics, community-shared open infrastructure for underwater applications, and integrated communications, navigation, and sensing. Dr. Song is a recipient of the National Science Foundation CAREER Award in 2021. He is an Associate Editor for Journal of Acoustical Society of America.
\end{IEEEbiography}

\end{document}